\documentclass[twocolumn, secnumarabic, amssymb, nobibnotes, aps, prb, 10pt, superscriptaddress]{revtex4-2}

\usepackage{lipsum} 

\usepackage{amsmath} 
\usepackage{mathtools}
\usepackage{bm} 
\usepackage{physics2}
\usephysicsmodule{ab} 
\usephysicsmodule{ab.braket}
\usephysicsmodule{diagmat}
\usephysicsmodule[showleft=2,showtop=2]{xmat} 
\usephysicsmodule{doubleprod}
\usepackage{derivative}
\usepackage{mathrsfs} 
\usepackage[dvipsnames]{xcolor}
\usepackage{siunitx}
\usepackage[version=4]{mhchem}
\usepackage[
    setpagesize=false,
    bookmarksopen=true,
    bookmarksnumbered=true,
    colorlinks=true, 
    linkcolor=OrangeRed,  
    citecolor=blue, 
    urlcolor=ForestGreen, 
    ]{hyperref}
\usepackage{xr}
\usepackage{footnotebackref}

\makeatletter
\newcommand\vb{\@ifstar\boldsymbol\mathbf}
\newcommand\va[1]{\@ifstar{\vec{#1}}{\vec{\mathrm{#1}}}}
\newcommand\vu[1]{%
\@ifstar{\hat{\boldsymbol{#1}}}{\hat{\mathbf{#1}}}}
\makeatother

\usepackage{amsfonts,amsmath,amssymb,amsthm}
\usepackage{graphicx}
\usepackage{bm}
\usepackage{color}
\usepackage{braket}

\usepackage{comment}

\usepackage{dcolumn}
\usepackage{ulem}
\usepackage{hyperref}

\begin{document}

\title{
Generation of magnetic chiral solitons, skyrmions, and hedgehogs with electric fields
}
\author{Teruya Nakagawara}
\email{nakagawara@gmail.com}
\affiliation{Department of Physics, Chiba University, Chiba 263-8522, Japan}
\author{Minoru Kanega}
\email{m.kanega.phys@chiba-u.jp}
\affiliation{Department of Physics, Chiba University, Chiba 263-8522, Japan}
\author{Shunsuke C. Furuya}
\email{furuya@saitama-med.ac.jp}
\affiliation{Department of Liberal Arts, Saitama Medical University, Moroyama, Saitama 350-0495, Japan}
\affiliation{Institute for Solid State Physics, The University of Tokyo, Kashiwa, 277-8581, Japan}
\author{Masahiro Sato}
\email{sato.phys@chiba-u.jp}
\affiliation{Department of Physics, Chiba University, Chiba 263-8522, Japan}

\begin{abstract}
Electric-field controls of Dzyaloshinskii-Moriya interactions (DMIs) have recently been discussed from the microscopic viewpoint. 
Since the DMI plays a critical role in generating topological spin textures (TSTs) such as the chiral soliton, the magnetic skyrmion, and the magnetic hedgehog, electric-field controls of these TSTs have become an important issue.
This paper shows that such electric-field-induced DMI indeed creates and annihilates TSTs by numerically solving the Landau-Lifshitz-Gilbert (LLG) equation for many-body spin systems at finite temperatures.
We show that when a strong electric field is applied in a proper way to one- or two-dimensional ferromagnets, the Hamiltonians are changed into the well-known spin models for the chiral soliton or the skyrmion lattice, and the TST states emerge. We utilize a machine-learning method to count the number of generated TSTs.
In the three-dimensional (3D) case, we demonstrate the electric-field induction of a magnetic hedgehog structure as follows: Applying a strong enough electric field along a proper direction to a skyrmion-string state (a triple-$\bm q$ state) at low but finite temperatures,
we find that the field-induced DMI can drive a quadruple-$\bm q$ state with hedgehog-antihedgehog pairs.
This result indicates that we have succeeded in constructing a simple 3D short-range interacting spin model hosting a magnetic hedgehog structure.
\end{abstract}

\date{\today}

\maketitle

\section{Introduction}

Recent technological advances enable us to apply stronger ac or dc electromagnetic fields to condensed matters than ever before,  making it possible to drive condensed matters into novel states that would not be realized without those strong fields.
The state-of-the-art pulsed laser technology gives the strong ac electric field and enlarges the possibility of the Floquet engineering of materials~\cite{Shirley_65,Sambe_73,Eckardt_15,Eckardt_17,Mikami_16,Sato_inbook_21,Oka_19}.

\begin{figure}[t!]
  \centering
  \includegraphics[width=\linewidth]{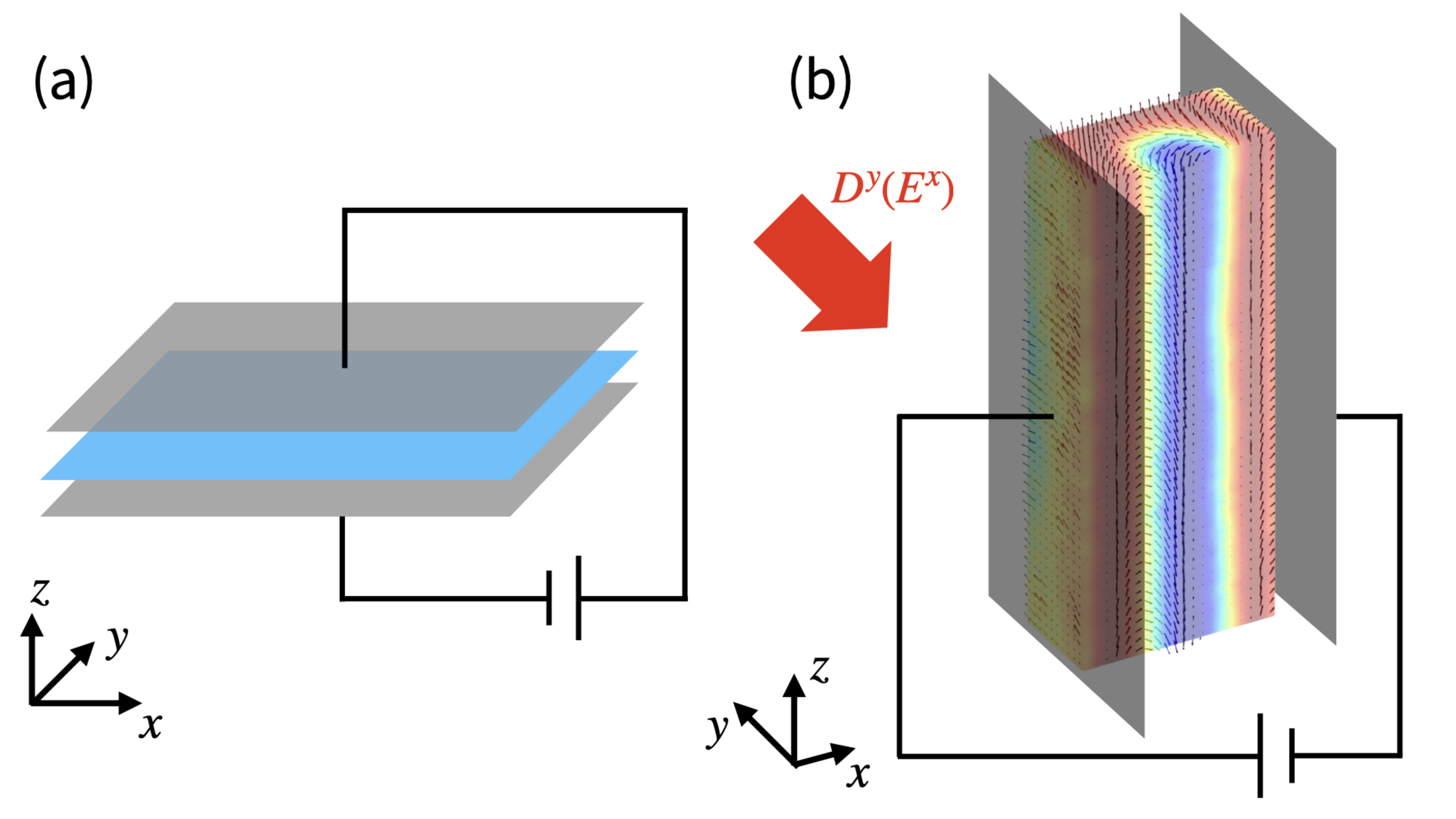}
  \caption{
  (a) Application of dc electric field to two-dimensional magnetic Mott-insulator film, where the soliton and skyrmion emerge.
  (b) Application of dc electric field to three-dimensional magnetic Mott insulator, where the magnetic hedgehog emerges.
  }
  \label{fig:condenser}
\end{figure}

Notably, dc electric fields remain advantageous to controlling materials from several practical viewpoints such as avoidability of an unwanted heating problem even under the prosperity of the Floquet engineering~\cite{Lazarides_heating_2014,D'alessio_heating_2014}.
Strong dc electric fields are available thanks to developments of \textit{e.g.}, field-effect transistors, scanning tunneling microscopes, and ferroelectric devices~\cite{Bisri_17,Ueno_14,Yazawa_21,Mikolajick_21,Chen_13,Braun_12,Kim_20,Huang_21,Magtoto_00} (Fig.~\ref{fig:condenser}).
The strength of those dc electric fields nowadays rises to $\sim 10~\mathrm{MV/cm}$~\cite{Yazawa_21,Mikolajick_21}.
Hence, dc electric-field controls of materials are a classic but cutting-edge approach to condensed-matter physics.

Some of the authors developed microscopic theories about dc electric-field controls of magnetism in Mott insulators~\cite{Takasan_19,Furuya_21,Furuya_24}.
Vast researches have been performed on dc electric-field controls of dielectrics, multiferroics~\cite{knb,mostovoy_FerroelectricitySpiralMagnets_2006,Tokura_14}, and spintronics~\cite{Matsukura_15}.
The series of theoretical papers~\cite{Takasan_19,Furuya_21,Furuya_24} are focused on how dc electric fields affect microscopically the many-body Hamiltonian of electrons in the Mott insulator and eventually its low-energy effective spin Hamiltonian.
Reference~\cite{Furuya_24} indicates how dc electric fields microscopically modulates the Dzyaloshinskii-Moriya interaction (DMI) of magnetic Mott insulators.
Such dc electric-field controls of DMI will enable us to create (or annihilate) a topological spin texture (TST) such as magnetic chiral solitons~\cite{Togawa_12,Kishine_15,Togawa_16}, magnaetic skyrmions~\cite{Muhlbauer_09,Yu_10,Seki_12,Kezsmarki_15,Kurumaji_2017}, and magnetic hedgehogs~\cite{Kanazawa_11,Tanigaki_15,Kanazawa_16,Fujishiro_19}.

TSTs have attracted attention for their topological stability and efficient energy consumption that match potential demands toward future device applications~\cite{Fert_13,Zhou_15,Zhang_15}.
For example, recent experimental studies revealed that we can apply strong dc electric fields of $ 1 \sim 10~\mathrm{MV/cm}$ to two-dimensional interfacial systems.
On the other hand, recent papers~\cite{Bogdanov_01,Yu_10,Seki_12,Seki_12_2,Sampaio_2013,Ryu_13} reported the realization of magnetic skyrmions in such interfaces of materials.
Moreover, electric-field controls of magnetic orders have also been under active investigations in atomic-layer materials~\cite{Huang2017_van-der-Waals,Jiang2018_van-der-Waals}. 
It is thus important to investigate how dc electric fields actually create TSTs based on the basic equation of motion and how stably those electric-field-driven TSTs can travel in the Mott insulator under the influence of spin-spin correlations and finite-temperature noise.
Temperature plays an essential role in transforming a trivial state into the TST state: 
TSTs are topologically stable, as their names suggest, and this means that there is an energy barrier between the TST state and the trivial one.
A finite temperature introduces an energy fluctuation and enables the trivial state to jump over the energy barrier and become the TST state.
In addition, all experiments are performed at finite temperatures.

This paper discusses dc electric-field controls of TSTs in magnetic Mott insulators based on numerical analyses of time evolution of TSTs under the influence of dc electric fields. We numerically solve the Landau-Lifshitz-Gilbert equation to investigate the microscopic spin dynamics with tuning dc electric fields. 
As already mentioned, the dc electric field yields a DMI in spin systems~\cite{Furuya_24}. 
We show that such an electric-field-induced DMI strongly affects the spin dynamics and resulting spin textures.
The key point of our study is to uncover the real-time evolution from a trivial state to a TST one.
Such an analysis of the time evolution under the microscopic equation of motion is yet to be investigated despite its importance.
Certainly, well-studied dynamical structure factors $S(\bm q,\omega)$ possess information about spin dynamics of the linear response. 
However, the transformation process from a trivial state to a TST one or vice versa is beyond the linear response. 
Therefore, witnessing the time evolution with tuning electric fields is worth performing because of its relevance to theories, experiments, and future applications.

We confirm that the electric-field-induced DMI indeed eventually turns the forced ferromagnetic state into TST states such as the chiral soliton lattice (CSL) state, the magnetic skyrmion lattice (SkX) state, and the quadruple-$\bm q$ state with hedgehog structures.
The last case is particularly interesting because no generic Hamiltonian with short-range interactions is yet available to realize such a magnetic hedgehog. 
Our numerical analyses indicate that we can construct a short-range interacting spin model hosting the magnetic hedgehog structure.

This paper is organized as follows.
Section~\ref{sec:method} describes our numerical method to investigate the spin dynamics accompanying changes of topological numbers.
While the framework presented in Sec.~\ref{sec:method} universally applies to the systems dealt with in this paper, details of the spin dynamics depends on a specific choice of the spin Hamiltonian.
Section~\ref{sec:hamiltonian} gives three Hamiltonians used in this paper.
This paper discuss the spin dynamics in three cases; one-dimensional (1D), two-dimensional (2D), and three-dimensional (3D) cases, where the dimensionality refers to the relevant spatial dimension of the lattice of the model.
We show the dynamical generation of the chiral soliton number in Sec.~\ref{sec:1d}, the skyrmion number in Sec.~\ref{sec:2d}, and the magnetic monopole charge in Sec.~\ref{sec:3d}.
We discuss characteristics of the spin dynamics and the generated topological number in each case.
The paper is summarized in Sec.~\ref{sec:summary}.

\section{Methods}\label{sec:method}

Our discussions stand on spin models of magnetic Mott insulators whose magnetic properties are described by Heisenberg-like models.
The Heisenberg model is an effective model that descibes low-energy physics of the Hubbard model at half filling.
However, to handle dc electric field effects in Heisenberg-like spin models properly, one needs to return to the Hubbard-like model of electrons even if one focuses on the low-energy physics.
The Heisenberg (super)exchange interaction results from direct hoppings of electrons between magnetic ions or indirect ones mediated by ligand ions~\cite{Goodenough_55,Kanamori_59}.
When the material has the Rashba spin-orbit coupling (SOC), the electric field yields spin-dependent electron hoppings and eventyally results in magnetic anisotropies, typically, the DMI~\cite{Furuya_24}.
The switching on (and off) of DMI by electric fields will be critical to electric-field controls of TST.
In fact, the presence of DMI is a key to realizing TST such as the CSL~\cite{Togawa_12,Kishine_15,Togawa_16} and the SkX~\cite{seki2016skyrmions}.
Reference~\cite{Furuya_24} suggests that dc electric fields can generate the CSL when applied to one-dimensional (1D) systems and the SkX when applied to two-dimensional (2D) ones on the square lattice at the Hamiltonian level.
As we already mentioned, generic Hamiltonian that realizes the magnetic hedgehog in three-dimensional (3D) systems is yet unavailable.
We show later how we create the magnetic hedgehog dynamically based on an equation of motion governed by the microscopic Hamiltonian.

\subsection{Time evolution based on LLG equation}

\begin{figure}[t!]
    \centering
    \includegraphics[width=\linewidth]{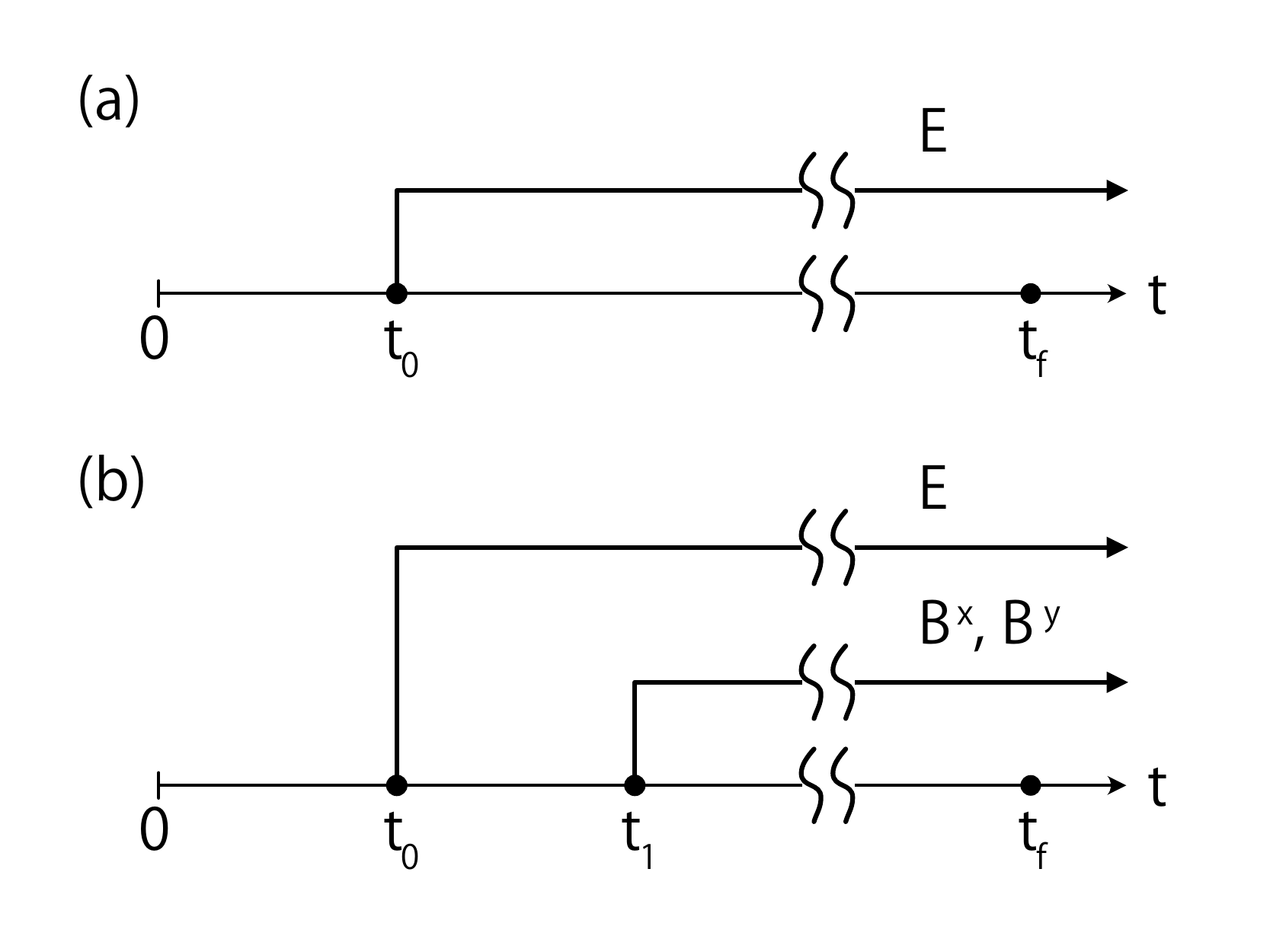}
    \caption{Schematic figure of important clock times in our spin dynamics simulations for (a) 1D and 2D cases and (b) 3D case.
    (a) We prepare an initial state at $t=0$ and make it evolve with time under the LLG equation for $t>0$. 
    Note that the initial state at $t=0$ is an equilibrium state of the intrinsic Hamiltonian at a given temperature $T$.
    When $t=t_0$, we switch the dc electric field $\bm E$ on and keep it on for $t>t_0$.
    When $t=t_f \gg t_0$, the system reaches a steady state.
    (b) We prepare an initial state at $t=0$ and make it evolve with time under the LLG equation for $t>0$.
    When $t=t_0$, we switch the dc electric field $\bm E$ and keep it on for $t>t_0$.
    When $t=t_1>t_0$, we also switch the magnetic fields $B^x$ and $B^y$ on and keep it on for $t>t_1$.
    When $t=t_f \gg t_1$, the system reaches a steady state.
    }
    \label{fig:protocol}
\end{figure}

Standing on the microscopic theory~\cite{Furuya_24}, we discuss how the microscopic equation of motion under the influence of dc electric fields actually creates TSTs such as CSL, SkX and the magnetic hedgehog using numerical techniques.
Let us describe our numerical methods before addressing details of models.
We adopt the Landau-Lifshitz-Gilbert (LLG) equation,
\begin{align}
    \frac{d\bm{S}_{\bm{r}}(t)}{dt}
    &= \frac{1}{\hbar} \bm{S}_{\bm{r}}(t) \times \bm{B}_{\mathrm{eff};\bm{r}}(t) +\frac{\alpha}{S}\bm{S}_{\bm{r}}(t)\times\frac{d\bm{S}_{\bm{r}}(t)}{dt},
    \label{LLG_eq_def}
\end{align}
where $\bm{S}_{\bm{r}}$ is the spin at the location $\bm{r}$, $\bm{B}_{\mathrm{eff};\bm{r}}(t)$ is an effective magnetic field that the spin $\bm{S}_{\bm{r}}$ feels, and $\alpha$ is the Gilbert damping coefficient. Each spin $\bm{S}_{\bm{r}}$ should be regared as an operator from the fully quantum mechanical viewpoint, but it is defined as a three-dimensional vector $(S^x_{\bm{r}},S^y_{\bm{r}},S^z_{\bm{r}})$ with length $S$ (i.e., spin quantum number) in the framework of the LLG equation. This semi-classical approximation is justified if the system is in or around magnetically ordered states, which we will consider in this study.
The dimensionless coefficient $\alpha$ depends on details of materials.
Typically, $\alpha=O(10^{-1})$ for metals and $\alpha=O(10^{-2})$ for insulators.
In this paper, we fix $\alpha=0.01$. 
The effective magnetic field $\bm{B}_{\mathrm{eff};\bm{r}}(t)=\partial\mathcal{H}/\partial\bm{S}_{\bm{r}}+\bm{\xi}_{\bm{r}}(t)$ depends on the spin Hamiltonian $\mathcal{H}$ and a random magnetic field $\bm{\xi}_{\bm{r}}(t)$.
We take into account effects of the temperature through the random field $\bm{\xi}_{\bm{r}}(t)=(\xi^x_{\bm{r}}(t),\,\xi^y_{\bm{r}}(t),\,\xi^z_{\bm{r}}(t))$ and suppose it as a white noise characterized by the following relations:
\begin{align}
    \braket{\xi_{\bm{r}}^\mu(t)} 
    &= 0, 
    \label{white_noise_av} \\
    \braket{\xi_{\bm{r}}^\mu(t)\xi_{\bm{r'}}^\nu(t')}
    &= \frac{2\hbar\alpha k_BT}{S} \delta_{\bm{r},\bm{r}'}\delta_{\mu,\nu}\delta(t-t').
    \label{white_noise_corr}
\end{align}
Here, $T$ is the temperature of the Mott insulator and $S$ is the spin quantum number to be set to $S=1$ in our numerical calculations. 
The deterministic part of the effective magnetic field is governed by the spin Hamiltonian $\mathcal{H}$, into which we plug a specific model Hamiltonian defined later.
We obtain time evolution of the spin operator by numerically solving the microscopic LLG equation based on Heun's method, as implemented in the \texttt{DifferentialEquations.jl} package~\cite{Rackauckas2017DifferentialEquationsjl}.

Let us describe protocols of how to apply electromagnetic fields to the spin system, schematically summarized in Fig.~\ref{fig:protocol}.
There are several important time scales $t_0$, $t_1$, and $t_f$.

For the 1D and 2D cases of Fig.~\ref{fig:protocol}~(a), we investigate the spin dynamics under dc electric fields as follows.
First, we prepare an initial state and set the clock time to $t=0$.
We choose the equilibrium state of the Hamiltonian at a given temperature $T$ as the initial state. Namely, we take a long enough time evolution (relaxation) by the LLG equation in the range $t<0$ to prepare the equilibrium state at $t=0$.
Note that the temperature $T$ is the same as that in the LLG equation thorugh the correlation \eqref{white_noise_corr} of the random field $\bm \xi_{\bm r}$.
For $t=t_0$, we switch on the dc electric field $\bm E$ and keep it on for $t>t_0$.
The dc electric field changes the spin dynamics by adding the DMI to the spin Hamiltonian $\mathcal H$.
We make the system evolve until it reaches a steady state, say, at $t=t_f$.
For instance, the exchange coupling $J=10~\mathrm{K}/k_B$ gives $t_f = 6000\hbar/J = 4583~\mathrm{ps}$. This time length is sufficiently larger than the typical time scale of spin dynamics in spin systems, $1-10^2$ ps.
We stop the time evolution at $t=t_f$ and obtain the steady state, which will naively be identical to equilibrium state.
However, this steady state is not always the same as the equilibrium state expected from the spin Hamiltonian at given electromagnetic fields and temperature.
We will come back to this point later in Sec.~\ref{sec:results}.

For the 3D case of Fig.~\ref{fig:protocol}~(b), we adopt a different, more complicated protocol. 
We first prepare an initial state and set the clock time to $t=0$.
We make it evolve for a while.
When $t=t_0$, we switch on the dc electric field and keep it on for $t>t_0$.
At a later time $t=t_1$ ($>t_0$), we switch on transverse magnetic fields $B^x$ and $B^y$ within the $xy$ plane. 
The reason for adding such a magnetic field is explained in Sec.~\ref{sec:hamiltonian} when we define the spin Hamiltonian.
Then, we make the system evolve with time until it reaches a steady state at $t=t_f$.
We take $t_f\gg t_1>t_0$.

It is important to obtain reliable statistical averages of physical quantities in each time step. 
Therefore, we execute the above protocols many times for each setup of temperature $T$, dc electromagnetic fields, and spin Hamiltonian.

\subsection{Topological number}

The numerical solution of the LLG equation \eqref{LLG_eq_def} gives us real-time information on the microscopic spin dynamics.
We can identify the creation and annihilation of TSTs from the time-evolving spin configuration by calculating a topological number $Q$ instantaneously at each time step.
For CSL in 1D lattice along the $x$ direction, 
$Q$ is the soliton number $Q_{\mathrm{CSL}}$ defined as
\begin{align}
    Q_{\mathrm{CSL}} = \frac{1}{2\pi} \int_\ell \mathrm{d}x\,\frac{\mathrm{d}\theta(x)}{\mathrm{d}x},
    \label{Q_CSL_def}
\end{align}
where the integral is performed over a spatial interval $\ell$ along the $x$ direction and $\theta(x)$ is an angle formed by two adjacent spins $\bm{S}_{x}$ at a position $x$ and $\bm{S}_{x+a_0}$ at an adjacent position $x+a_0$. 
Here, $a_0$ is the lattice spacing of the 1D spin chain.
A chiral soliton in the interval $\ell$ gives $Q_{\mathrm{QSL}}=\pm1$.

For SkX in 2D lattice on the $x$-$y$ plane, 
we use the skyrmion number $Q_{\mathrm{SkX}}$:
\begin{align}
    Q_{\mathrm{SkX}} = \frac{1}{4\pi}\int_A \mathrm{d}^2\bm{r}\, \frac{\partial \bm{n}_{\bm{r}}}{\partial x} \times \frac{\partial \bm{n}_{\bm{r}}}{\partial y} \cdot \bm{n}_{\bm{r}}.
    \label{Q_SkX_def}
\end{align}
Here, $\bm r= (x,y)$ denotes the position of the spin $\bm S_{\bm r}$ and $\bm{n}_{\bm{r}}=\bm{S}_{\bm{r}}/|\bm{S}_{\bm{r}}|$ is the unit vector parallel to $\bm{S}_{\bm{r}}$.
One magnetic skyrmion in the area $A$ gives an integer $Q_{\mathrm{SkX}}\in\mathbb{Z}$, typically, $Q_{\mathrm{SkX}}=\pm1$.
Note that we need to approximate spins $\bm{S}_{\bm r}$ as a continuous function of the space when we use the formulas of Eqs.~\eqref{Q_CSL_def} and \eqref{Q_SkX_def}. In other words, Eqs.~\eqref{Q_CSL_def} and \eqref{Q_SkX_def} are violated when the directions of neighboring spins are quite different from each other. 

For the magnetic hedgehog, we use the monopole charge $Q_m$ defined on each unit cell of a three-dimensional (3D) lattice.
A monopole (an antimonopole) gives $Q_m = 1$ ($Q_m=-1$, respectively) and otherwise $Q_m=0$.
For instance, in the case of a cubic lattice, the monopole charge of a monopole whose center is at $\bm{r}_c$ is
\begin{align}
    Q_m(\bm{r}_c) = \frac{1}{4\pi}\sum_{\mu=x,y,z}\sum_{\delta_\mu=\pm 1} \delta_\mu \Omega^\mu \biggl(\bm{r}_c + \frac{\delta_\mu}{2}\bm{e}_\mu\biggr), \label{Qm_hdg_def}
\end{align}
where $\bm{e}_\mu$ is the unit vector along the $\mu=x,y,z$ axis and $\delta_\mu=\pm 1$. The right-hand side in Eq.~\eqref{Qm_hdg_def} is defined as follows: 
The quantity $\Omega^\mu(\bm{r})$ resides on a plaquette normal to $\bm{e}_\mu$ at the location $\bm{r}$.
The plaquette formed by four spins is dividable into two triangles, $i=1$ and $i=2$, each of which is formed by three spins.
$\Omega^\mu(\bm{r})$ is the sum of solid angles over the $i=1$ and $i=2$ triangles: $\Omega^\mu(\bm{r})=\sum_{i=1,2}\Omega^\mu_i(\bm{r})$.
The three spins, say $\bm{S}_1$, $\bm{S}_2$, and $\bm{S}_3$, of the $i$th triangle gives the following solid angle~\cite{Yang_16,Okumura_20,Okumura_20_JPS,Shimizu_21,Shimizu_22}.
\begin{align}
  \Omega_i^\mu(\bm{r}) 
  = 2\tan^{-1}
  \ab( \frac{\bm{S}_1\cdot \bm{S}_2\times\bm{S}_3}{S_1 S_2 S_3 + \sum_{\mathrm{cyc\ab(1,2,3)}} \ab(\bm{S}_i\cdot\bm{S}_j)S_k}),
  \label{solid_angle_triangle_def}
\end{align}
where $ \sum_{\mathrm{cyc\ab(1,2,3)}}$ means the summation about $(i,j,k)=(1,2,3)$ and all the other cyclic transformations.

\subsection{Machine learning for counting topological numbers}\label{sec:ML}

To analyze properties of TSTs, we count the number of TSTs and then take a thermal average. 
Let us briefly explain a practical method to count the soliton number \eqref{Q_CSL_def} and the skyrmion number \eqref{Q_SkX_def} in our analyses.
We judge the creation and annihilation of those TSTs by using a combination of image-processing and machine-learning (ML) methods rather than by using the above-mentioned mathematical definitions.
We trained our detection model by giving it snapshots of TSTs as images in advance and adopting them in the 1D and 2D cases.

The ML-based method is built as follows.
Our method stands on an object-detection algorithm YOLO (You Only Look Once)~\cite{Redmon2016You}. 
We utilize an officially available pre-trained model, \texttt{YOLO11l}~\cite{Khanam2024YOLOv11}, based on the transfer learning.
In the 1D case, we trained the ML model for 234 epochs on 179 snapshots of chiral solitons,
where one epoch refers to a unit of 
learning through the entire training dataset (i.e., the 179 snapshots).
In the 2D case, we trained the ML model for 430 epochs for 410 snapshots of skyrmions.
We used the default hyperparameter settings provided by the \texttt{ultralytics YOLO11} package to train our detection model, without further tuning.
To avoid false positives of those topological spin textures as much as possible, we set a confidence score.
Any detection with the confidence score lower than 0.5 is to be automatically filtered out.

Here, we explain why we adopt the ML-based method instead of the topological numbers of Eqs.~\eqref{Q_CSL_def} and \eqref{Q_SkX_def}. 
The soliton number \eqref{Q_CSL_def} and the skyrmion number \eqref{Q_SkX_def} are defined
as integrals over an interval $\ell$ or an area $A$.
Proper choice of the domain ($\ell$ and $A$) is a priori unclear.
We have two options.
The first option is to choose the entire system itself as the domain.
In this choice, however,  serever boundary effects on the topological numbers are inevitable.
We adopt the open boundary condition in the 1D and 2D cases for physical and technical reasons.
Physically, the open boundary condition is realistic.
Suppose a situation that we put the thin film inside a capacitor [Fig.~\ref{fig:condenser}~(a)] to apply the dc electric field.
Technically, the open boundary condition is much more compatible with incommensurate configurations such as TST than the periodic boundary condition.
We thus do not take this option.

The second option is to choose the domain by looking at the TST so that it contains the TST.
Manual counting by eye quickly becomes impractical for large-scale simulations
Employing the ML-based method becomes more effective as the system size becomes larger or the simulation time becomes longer.
The ML-based method allows us to specify the domain where each TST exists.
We can confirm the accuracy of the ML-based detection by calculating the integrals \eqref{Q_CSL_def} and \eqref{Q_SkX_def}, where we take the domains $\ell$ and $A$ as those determined by the ML-based method.

We will show later how the ML-based detection method of TSTs works in Secs.~\ref{sec:1d} and \ref{sec:2d}.
Note that in the 3D case, we use the mathematical definition \eqref{Qm_hdg_def} for numerical detection of the magnetic hedgehog instead of the ML-based method since Eq.~\eqref{Qm_hdg_def} does not contain the ambiguity of the domain unlike Eqs.~\eqref{Q_CSL_def} and \eqref{Q_SkX_def}.

\section{Hamiltonians}\label{sec:hamiltonian}

\begin{figure}[t!]
  \centering
  \includegraphics[width=\linewidth]{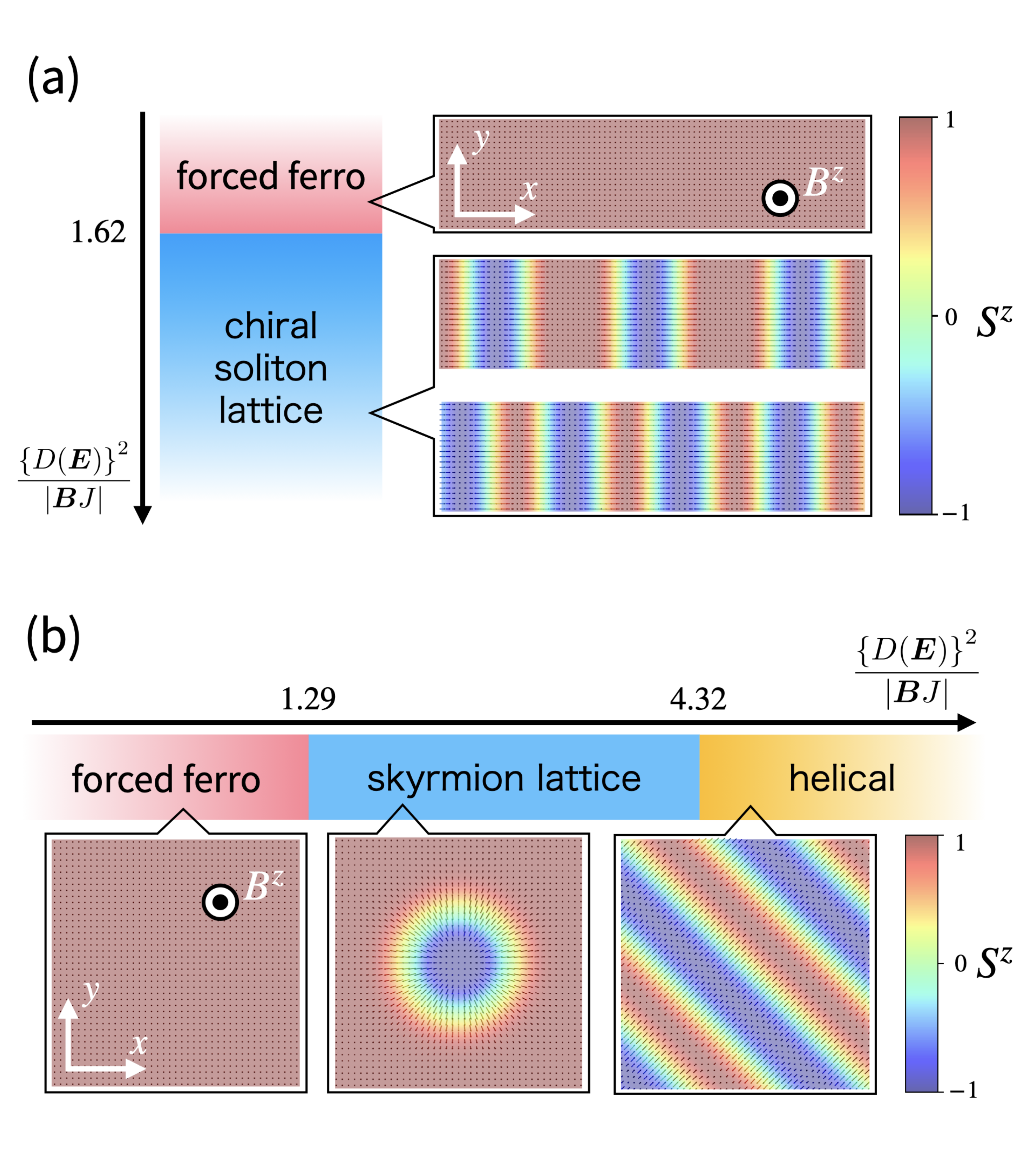}
  \caption{
  (a) Ground-state phase diagram of 1D spin chain \eqref{H_CSL_def}.
  The chiral soliton lattice (CSL) phase appears for the large DMI strength $D^2/|\bm B J|>1.62$~\cite{Kishine_15}.
  (b) Ground-state phase diagram of 2D spin model on square lattice \eqref{H_SkX_def}.
  The magnetic skyrmion lattice (SkX) phase appears for moderate DMI strength $1.29<D^2/|\bm BJ|<4.32$~\cite{seki2016skyrmions}.
  }
  \label{fig:phaseDiagram_CSL_SkX}
\end{figure}

This section is devoted to the definition and explanation about our spin systems. 
The spin Hamiltonian affects the electron-spin dynamics through $\bm{B}_{\mathrm{eff};\bm{r}}$.
Here we consider three model Hamiltonians for CSL, SkX, and the magnetic hedgehog.
This paper hereafter refers to these three models as the 1D, 2D, and 3D cases, respectively, following the dimensionality of the underlying lattice.

We discuss CSL in a 1D spin chain along the $x$ axis with the following Hamiltonian~\cite{Kishine_15}
\begin{align}
    \mathcal{H}_{\mathrm{CSL}}
    &=-J\sum_j \bm{S}_j\cdot\bm{S}_{j+1}+D(E^z)\sum_j\bm{e}_y\cdot\bm{S}_j\times\bm{S}_{j+1}
    \notag\\
    &\qquad -B^z\sum_jS_j^z,
    \label{H_CSL_def}
\end{align}
where $-J<0$ is the ferromagnetic exchange strength, $D(E^z)$ is the electric-field-induced DMI coupling, and $B^z$ is the magnetic field. Here, $\bm E$ is the external dc electric field and  
we suppose $D(0)=0$ in Eq.~\eqref{H_CSL_def}. 
The electric field turns on the DMI and drives the ferromagnetic state at $\bm{E}=E^z\bm e_z=0$ into the CSL state.
Reference~\cite{Furuya_24} gives detailed derivation of the Hamiltonian \eqref{H_CSL_def} under dc electric fields for an $S=1/2$ situation: the CuO$_2$ chain.
Cu ions are aligned along the $x$ direction whose nearest-neighbor ferromagnetic exchange interaction is mediated by O ions on the $xy$ plane.
Then, dc electric field needs to be along the $z$ direction implied in the Hamiltonian \eqref{H_CSL_def}. 
Within a simple estimation~\cite{Furuya_24}, we expect that the magnitude of the DMI, $D/J$, can approach to ${\cal O}(10^{-1})$ when a strong dc electric field of $\sim 1-10$ MV/cm is applied to proper Mott insulators.

The SkX phase emerges in a spin model on the 2D square lattice with the following Hamiltonian~\cite{seki2016skyrmions}.
\begin{align}
    \mathcal{H}_{\mathrm{SkX}}
    &= - J \sum_{\bm{r}} (\bm{S}_{\bm{r}}\cdot \bm{S}_{\bm{r}+\bm{e}_x} + \bm{S}_{\bm{r}}\cdot \bm{S}_{\bm{r}+\bm{e}_y})- B^z\sum_{\bm{r}}S_{\bm{r}}^z
    \notag \\
    &\quad +D(E^z) \sum_{\bm{r}}(
    \bm{e}_y \cdot \bm{S}_{\bm{r}}\times\bm{S}_{\bm{r}+\bm{e}_x} -\bm{e}_x\cdot\bm{S}_{\bm{r}}\times\bm{S}_{\bm{r}+\bm{e}_y}
    ).
    \label{H_SkX_def}
\end{align}
We put the square lattice on the $x$-$y$ plane [see Fig.~\ref{fig:phaseDiagram_CSL_SkX}~(b)].
The Dzyaloshinskii-Moriya (DM) vector of the DMI points to a normal direction to the bond~\cite{Furuya_24}, for instance, the vector $\bm{e}_y$ for the $\bm{e}_x$ bond between $\bm{r}$ and $\bm{r}+\bm{e}_x$. 
As a result, the DMI-driven skyrmion is of N\'eel type (not Bloch type).
When $\bm{E}=E^z\bm{e}_z=0$, the Hamiltonian \eqref{H_SkX_def} has the ferromagnetically ordered phase at low temperatures because of $D(0)=0$.
A dc electric field, making $D/J\not=0$, drives the ferromagnetic phase into the SkX phase~\cite{Furuya_24}.

Last but not least, let us define the spin Hamiltonian in the 3D case to discuss the electric-field induction of the magnetic hedgehog.
Note that a realistic Hamiltonian that gives the magnetic hedgehog as the thermodynamic phase is yet unknown.
Here, we consider a spin model on the cubic lattice whose Hamiltonian composed of three parts:
\begin{align}
    \mathcal{H}_{\mathrm{hdg}} = \mathcal{H}_{\mathrm{hdg}}^{\mathrm{ini}} + \mathcal{H}_{\mathrm{hdg}}^{\mathrm{ex1}}+
    \mathcal{H}_{\mathrm{hdg}}^{\mathrm{ex2}},
    \label{H_hdg_def}
\end{align}
where $\mathcal{H}_{\mathrm{hdg}}^{\mathrm{ini}}$ is the Hamiltonian in the absence of additional electric and magnetic fields, 
$\bm{E}$ and $\bm{B}$, which are respectively introduced in the time ranges $t>t_0$ and $t>t_1$ [see Fig.~\ref{fig:protocol}~(b)].
Switching $\bm{E}$ on at $t=t_0$ adds the extra interaction $\mathcal{H}_{\mathrm{hdg}}^{\mathrm{ex1}}$ to the Hamiltonian \eqref{H_hdg_def}. Furthermore, $\mathcal{H}_{\mathrm{hdg}}^{\mathrm{ex2}}$ appears at $t=t_1$ 
by introducing an additional magnetic field $\bm B$.

The initial Hamiltonian $\mathcal{H}_{\mathrm{hdg}}^{\mathrm{ini}}$ with $\bm{E}=\bm{B}=\bm 0$ is given by
\begin{align}
  \mathcal{H}_{\mathrm{hdg}}^{\mathrm{ini}}
  &= - J \sum_{\bm{r},\bm{\delta}}  \bm{S}_{\bm{r}} \cdot \bm{S}_{\bm{r}+\bm{\delta}}- B_0^z \sum_{\bm{r}} S^z_{\bm{r}} \notag \\
  &\quad - D^{xy} \sum_{\bm{r}} \ab[ \ab( \bm{S}_{\bm{r}} \times \bm{S}_{\bm{r}+\bm{e}_x} ) \cdot \bm{e}_x + \ab( \bm{S}_{\bm{r}} \times \bm{S}_{\bm{r}+\bm{e}_y} ) \cdot \bm{e}_y ],
  \label{eq:Hedgehog_ini}
\end{align}
where $\bm{\delta}=\bm{e}_x$, $\bm{e}_y$, $\bm{e}_z$ are the unit vectors that define the cubic lattice, $-J<0$ is the ferromagnetic exchange interaction, and $D^{xy}$ is an intrinsic DMI that has nothing to do with the external electric field $\bm{E}$.
Though the electric field can modify both $J$ and $D^{xy}$, we ignore this quantitative modification because it is less significant than the qualitative change $\mathcal{H}_{\mathrm{hdg}}^{\mathrm{ex1}}$ defined below. 
The magnetic field $B^z_0$ is applied from the beginning $t<0$, and 
the equilibrium state for Eq.~\eqref{eq:Hedgehog_ini} is realized at $t=0$. In the ground state of Eq.~\eqref{eq:Hedgehog_ini}, a SkX string state shown in Fig.~\ref{fig:condenser}~(b) is expected to appear because the model~\eqref{eq:Hedgehog_ini} is regarded as the layered system of ferromagnetically-coupled 2D spin Hamiltonians hosting a SkX phase: Each layer has a SkX state for proper values of $D^{xy}$ and $B_0^z$. In fact, we will start a SkX string state at $t=0$ in the numerical calculation of the time evolution. Note that the resulting SkX is of Bloch type because the DM vector is parallel to the bond with DMI in each 2D layer.

We consider a situation that the electric field $E^x$ for $t>t_0$ yields the following DMI of $\mathcal{H}_{\mathrm{hdg}}^{\mathrm{ex1}}$:
\begin{align}
    \mathcal{H}_{\mathrm{hdg}}^{\mathrm{ex1}}
    & = D^y\ab(E^x)\bm{e}_y \cdot \sum_{\bm{r}} \ab( \bm{S}_{\bm{r}} \times \bm{S}_{\bm{r}+\bm{e}_{{z}}} ). \label{eq:Hedgehog_ex}
\end{align}
We note that the above electric-field induced DMI on the $z$ bond, whose DM vector points to the $y$ direction, can appear in the spin system of Eq.~\eqref{H_hdg_def} if the corresponding Hubbard-like electron system possesses an electric-field induced Rashba SOC~\cite{Furuya_24}. The dc field $\bm E$ breaks the inversion symmetry and, as a result, the above DMI possibly appears. 
We apply the periodic boundary condition along the $y$ and $z$ directions, but apply the open boundary condition along the $x$ direction to make it compatible with the dc electric field $E^x$ along the $x$ direction [see Fig.~\ref{fig:condenser}~(b)].

In addition to $\mathcal{H}_{\mathrm{hdg}}^{\mathrm{ex1}}$, we introduce the following Zeeman interaction for $t\ge t_1$ [see Fig.~\ref{fig:protocol}~(b)] by applying additional magnetic field $\bm B$:
\begin{align}
\mathcal{H}_{\mathrm{hdg}}^{\mathrm{ex2}}=
- \sum_{\substack{\bm{r}\\\mu=x,y}} B^\mu S_{\bm{r}}^{\mu}. \label{eq:Hedgehog_ex2}
\end{align}
The point is that the $x$ component of the magnetic field in Eq.~\eqref{eq:Hedgehog_ex2} is perpendicular to the electric-field induced DM vector $D^y(E^x){\bm e}_y$:
\begin{align}
    \bm{B}=B^x\bm{e}_x+B^y\bm{e}_y.
    \label{B_transverse_def}
\end{align}

Finally, we explain why we adopt $\mathcal{H}_{\mathrm{hdg}}^{\mathrm{ex1}}$ and $\mathcal{H}_{\mathrm{hdg}}^{\mathrm{ex2}}$ as the external electromagnetic-field induced Hamiltonian. 
As we mentioned, for $0\le t < t_0$, an equilibrium SkX string state [Fig.~\ref{fig:condenser}~(b)] is realized due to the Hamiltonian $\mathcal{H}_{\mathrm{hdg}}^{\mathrm{ini}}$. 
The SkX string state (and SkX phase in the 2D model) 
can be viewed as a triple-$\bm q$ state, in which three $\bm q$ vectors, three magnetic ordering vectors, are in the $x$-$y$ plane. 
If we can add a spiral spin structure along the $z$ direction perpendicular to the $x$-$y$ plane, a quadrupole-$\bm q$ state is expected to emerge and such a state is known to possess a spin texture with hedgehog-antihedgehog pairs~\cite{Yang_16,Okumura_20,Okumura_20_JPS,Shimizu_21,Shimizu_22,Kanazawa_11,Tanigaki_15,Kanazawa_16,Fujishiro_19}.
This is the reason why we consider the electric field $\bm E$ along the $x$ direction and the resulting DMI on the $z$ bond. 
The importance of the magnetic field $\bm B$ is understood from the 1D and 2D spin systems with TSTs. 
In both 1D and 2D models of Eqs.~\eqref{H_CSL_def} and \eqref{H_SkX_def}, the DM vector is perpendicular to the magnetic field, and the competition between the DMI and the Zeeman interaction stabilizes the TSTs. Therefore, even in the 3D case, if a magnetic field perpendicular to $D^y\ab(E^x)\bm{e}_y$ in Eq.~\eqref{eq:Hedgehog_ex} is introduced, a quadrupole-$\bm q$ state with hedgehogs is expected to be more stabilized. 
Therefore, we introduce additional Zeeman interaction $\mathcal{H}_{\mathrm{hdg}}^{\mathrm{ex2}}$ together with the DMI $\mathcal{H}_{\mathrm{hdg}}^{\mathrm{ex1}}$. The key point is that the magnetic field $B^x$ is perpendicular to 
the electric-field induced DM vector $D^y\ab(E^x)\bm{e}_y$. 
We will show that the above strategy of applying $\bm E$ and $\bm B$ works well in the next section.

\section{Results}\label{sec:results}

Let us move on to analyses of numerical results obtained from the LLG equation \eqref{LLG_eq_def} with the Hamiltonian \eqref{H_CSL_def} for the 1D case, \eqref{H_SkX_def} for 2D, and \eqref{H_hdg_def} for 3D.
As already mentioned, we prepare an initial equilibrium state at $t=0$ and make it evolve along with the LLG equation.
For the sake of generating TSTs by applying electromagnetic fields, it is important to choose convenient and realistic initial states.
We adopt almost all-up (forced ferromagnetic) equilibrium states at a finite temperature as the initial state in the 1D and 2D cases. 
In the 2D case, we also consider another choice of the initial state for comparison, as we discuss later in Sec.~\ref{sec:2d}.
Our choice of the initial state in the 3D case is slightly complicated. 
As we mentioned in the previous section, we take a SkX-string state as the initial equilibrium state at $t=0$. That is, we start the time evolution from a topological state with a skyrmion string, differently from the trivial ferromagnetic states in the 1D and 2D cases. 

Then we show that the SkX-string state turns into a magnetic hedgehog state with magnetic monopoles as a result of the time evolution under the dc electric field. 
In any case, we prepare an initial state and make it evolve along with the LLG equation under the influence of dc electric fields.

The LLG equation contains the random field $\bm \xi(t)$ that incorporates the finite-temperature effect.
Thanks to this effect, the steady state obtained after a long enough time evolution can effectively be a thermal equilibrium state at the given temperature. (We note that as we will see soon later, the steady state cannot approach the equilibrium state in some cases.)
The random field follows the canonical distribution represented by Eqs.~\eqref{white_noise_av} and \eqref{white_noise_corr}.
To evaluate topological charges at a temperature $T$, we need to take the thermal average, namely, the average of steady states with respect to the canonical distribution of the random field.
Note that taking the average of the topological charge over those steady states generated by the finite-temperature LLG equation is equivalent to taking the thermal average.
We take the average over $N_s=O(10^2)$ steady states in the analyses throughout this paper.

In the 1D and 2D cases, we apply the dc electric field $\bm{E}(t)=E^z\bm{e}_z\theta(t-t_0)$ and the dc magnetic field $\bm{B}=B^z\bm{e}_z$, where $\theta(z)$ is the Heaviside step function. 
In the 3D case, we apply $\bm{E}(t)=E^y\bm{e}_y\theta(t-t_0)$ and $\bm{B}(t)=(B^x\bm e_x+B^y\bm e_y)\theta(t-t_1)$.
We again note taht there are two kinds of magnetic fields, $B^z$ in the initial Hamiltonian \eqref{eq:Hedgehog_ini} and $\bm B$ in the externally induced interaction \eqref{eq:Hedgehog_ex2}.
The above-mentioned $\bm B(t)$ refers to the latter transverse magnetic field.

Despite the time dependence of $\bm{E}(t)$, we call it as the dc field because the strength is kept constant once it is switched on at $t=t_0$.
For $t<t_0$, since we have $\bm{E}(t)=0$, the system is in the phase that the the ground state of the Hamiltonian without $\bm{E}$ belongs to.
The ground state is the forced ferromagnetic state in the 1D and 2D cases and the SkX-string state in the 3D case.
The electric field $\bm{E}(t)$ is turned on at $t=t_0$ and keeps its strength constant for $t>t_0$.
In what follows, we show snapshots of the spin configurations evolving in accordance with the LLG equation \eqref{LLG_eq_def} under the influence of dc electric fields.
Note that we call the spin configuration at a given fixed time $t$ the snapshot in the remainder of the paper.

\subsection{1D: Chiral soliton lattice}\label{sec:1d}

\begin{figure}[t!]
  \centering
  \includegraphics[width=\linewidth]{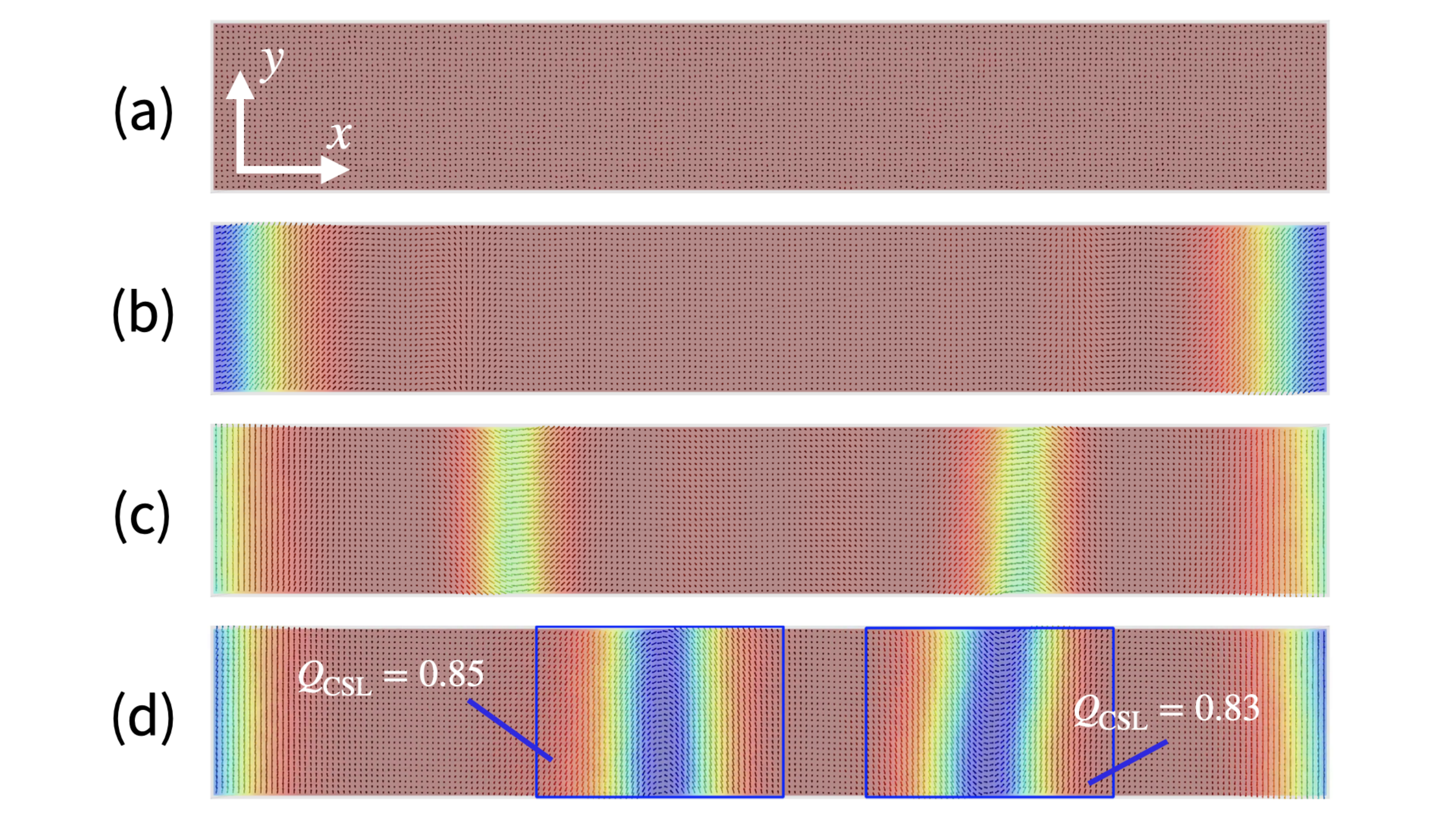}
  \caption{
  Time evolution of the spin chain~\eqref{H_CSL_def} with $D=0.18J$.
  The system size is set to $(N_x, N_y) = (200,30)$.
  We show how the dc electric field induces chiral solitons by looking at snapshots of the time evolution of the LLG equation.
  The four panels show (a) the forced-ferromagnetic state as the initial state at $t=0$, (b) the state at $t=200\hbar/J$, (c) the state at $t=1.0\times 10^3\hbar/J$, and 
  (d) the state at $t=6.0\times 10^3\hbar/J$, respectively. 
  We detect the chiral solitons in the panels (c) and (d) by using the method combining the image processing and the machine learning.
  Solitons are detected in the two regions indicated by the blue boxes. 
  The number [Eq.~\eqref{Q_CSL_def}] attached to each box shows the soliton number of the spin texture inside the box, indicating the existence of the chiral soliton therein.
  We follow the motion of the chiral soliton in the course of the time evolution (see Movie-1 in Supplementary Material~\cite{suppl}). 
  }
  \label{fig:CSL_dynamics}
\end{figure}

Figure~\ref{fig:CSL_dynamics} shows how the electric-field-induced DMI drives the initial ferromagnetic state into the CSL state.
We set parameters of $\mathcal{H}_{\mathrm{CSL}}$~\eqref{H_CSL_def} to $J=1$, $\alpha=0.01$, $k_BT/J=10^{-3}$ (a low but non-zero temperature), and $B^z/J=0.01$.
We prepare the 1D spin chains on the strip in the $x$-$y$ plane. The length for the $x$ ($y$) direction is set to $N_x=200$ ($N_y=30$) sites. 
Along the $x$ direction, the spin system has the nearest-neighbor ferromagnetic exchange interaction and the uniform DMI as given by Eq.~\eqref{H_CSL_def}. 
On the other hand, we assume that along the $y$ direction of Fig.~\ref{fig:CSL_dynamics}, the system has a simple nearest-neighbor ferromagnetic exchange interaction $-J$.
We take the $N_y>1$ width since the experimental realization of the 1D spin chain is not pure 1D but quasi-1D. 
We impose the open boundary condition on both $x$ and $y$ directions, and the condition generally makes the generation of incommensurate spin textures (chiral solitons, skyrmions, etc.) easier than the periodic boundary condition.

\begin{figure}[t!]
  \centering
  \includegraphics[width=\linewidth]{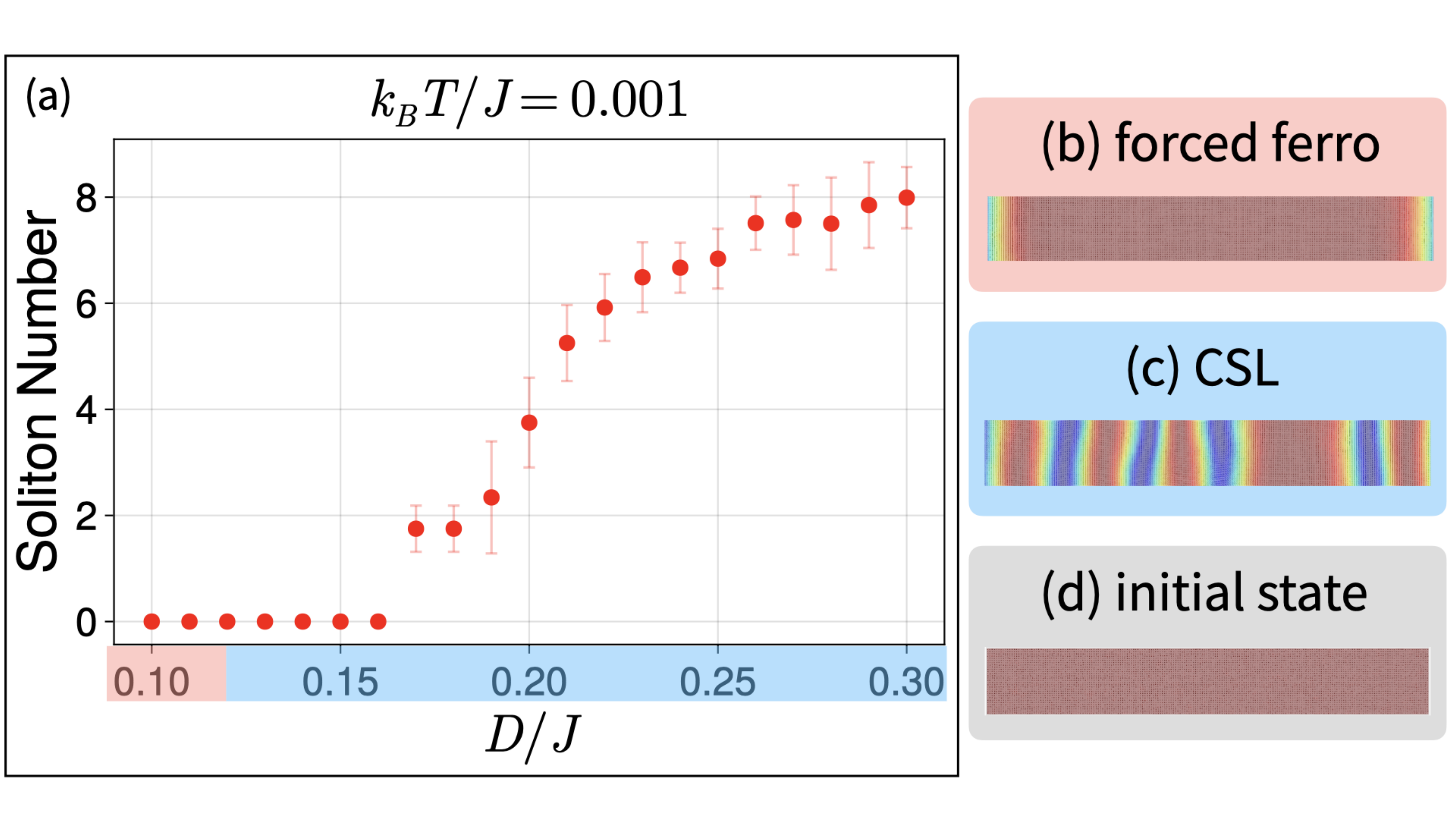}
  \caption{
  (a) $D/J$ dependence of the soliton number \eqref{Q_CSL_def} in steady states at $t=t_f$ and  temperature $k_BT/J=0.001$.
  (b) Typical spin configuration of the forced ferromagnetic steady state for $D/J=0.1$ obtained at $t=t_f$ after the long-enough time evolution.
  (c) Typical spin configuration of the CSL steady state for $D/J=0.20$. 
  (d) Initial state used for time evolution. 
  Each color on the horizontal axis in panel (a) represents the ground-state phase: Red (blue) color corresponds to the forced ferromagnetic (CSL) phase.
  }
  \label{fig:CSL_average}
\end{figure}

Figures~\ref{fig:CSL_dynamics}~(a) and (b) show the ferromagnetic state as the initial state at $t=0$ and the state at $t=2t_0$, respectively.
On the other hand, Fig.~\ref{fig:CSL_dynamics}~(c) and (d) show states at $t$ with $t_0\ll t < t_f$ and $t=t_f$, respectively.
These figures demonstrate that the initial ferromagnetic state at $t=0$ turns into the CSL state in the case of $D(E^z)^2/JB^z= 2.86$.
The dc electric field affects the spin dynamics through the DMI strength $D(\bm E)$ in the Hamiltonian \eqref{H_CSL_def}.
Figures~\ref{fig:CSL_average}~(a) shows how the soliton number $Q_{\mathrm{CSL}}$ [Eq.~\eqref{Q_CSL_def}] of the steady state at $t=t_f$ depends on the ratio $D(\bm{E})/J$ with fixed $|\bm B|/J=0.01$. 
The state remains in the forced ferromagnetic phase for small $D(\bm{E})/J$ even after the long-time evolution, while it enters the CSL phase for large $D(\bm{E})/J$.
The soliton number grows with the increase of $D(\bm{E})/J$.

The range $D(\bm{E})/J\le 0.16$ where we found $Q_{\mathrm{CSL}}=0$ in Fig.~\ref{fig:CSL_average}~(a) is slightly wider than the known range of the forced-ferromagnetic phase at $T=0$~\cite{Kishine_15},
\begin{align}
    0\le D/J\le D^c = \frac 4{\pi}\sqrt{B_zJ}\approx 0.13
    \label{critical_DMI_strength}
\end{align}
estimated from the phase diagram [see also Fig.~\ref{fig:phaseDiagram_CSL_SkX}~(a)] indicated as the red bar on the horizontal axis. 
The derivation of the critical value $D^c$ separating the forced-ferromagnetic and the CSL phases at $T=0$ is reviewed in App.~\ref{app:Dc}.
We believe that the following finite-size effect is one of the main reason why this overestimation of the critical $D^c$ occurs.
The CSL is a lattice structure of chiral solitons aligned with a certain distance.
The larger the DMI strength becomes, the shorter the lattice spacing between neighboring solitons becomes.
This distance diverges as the ratio $D/J$ approaches the critical value that separates the forced-ferromagnetic and the CSL phases.
Hence, when $D(E^z)/J$ is quite small, the length of the finite-size system may be shorter than the distance between neighboring solitons and the finite-size system cannot possess chiral soliton.
This unfortunate missing of the chiral soliton in the finite-size system would be largely responsible for the fact that our numerical simulation overestimates the critical value of $D/J$ between the forced-ferromagnetic and the CSL phases.

A few important differences between the true ground state and the steady state at $t=t_f$ also emerge due to temperature. 
We discuss this temperature effect on the creation of the chiral soliton.
As we mentioned earlier, we need the finite temperature to turn the trivial forced-ferromagnetic state into the CSL state. 
Namely, a large enough thermal fluctuation is necessary to go beyond the energy barrier between trivial and non-trivial TST states. 
However, at the same time, finite temperature also exhibits the opposite effect of destroying the chiral soliton.
This destabilization by finite temperatures is generally manifest in low-dimensional (1D and 2D) systems.
We thus expect a moderate temperature range to maximize the creation of the chiral soliton. 
The creation of the chiral soliton will be suppressed when the temperature is far from the moderate range. 
We also comment on finite-temperature fluctuations of the CSL state. 
At zero temperature, the chiral soliton forms the lattice, as the name of CSL suggests.
However, at finite temperatures, the chiral solitons fluctuate and hardly form the lattice. As a result, we observe a sort of soliton-gas state (see Appendix D~\ref{app:movie_1d} for a movie of time evolution).

In this subsection, we showed results of the spin dynamics governed by the LLG equation under the influence of dc electric-field-induced DMI in the 1D case.
We confirmed that the long-enough time evolution indeed drives the spin chain into the CSL state as shown in Fig.~\ref{fig:CSL_dynamics}.
In addition, the ML-based method turned out to be effective as a detection method of chiral solitons.

\subsection{2D: Skyrmion lattice}\label{sec:2d}

\begin{figure}[t!]
  \centering
  \includegraphics[width=\linewidth]{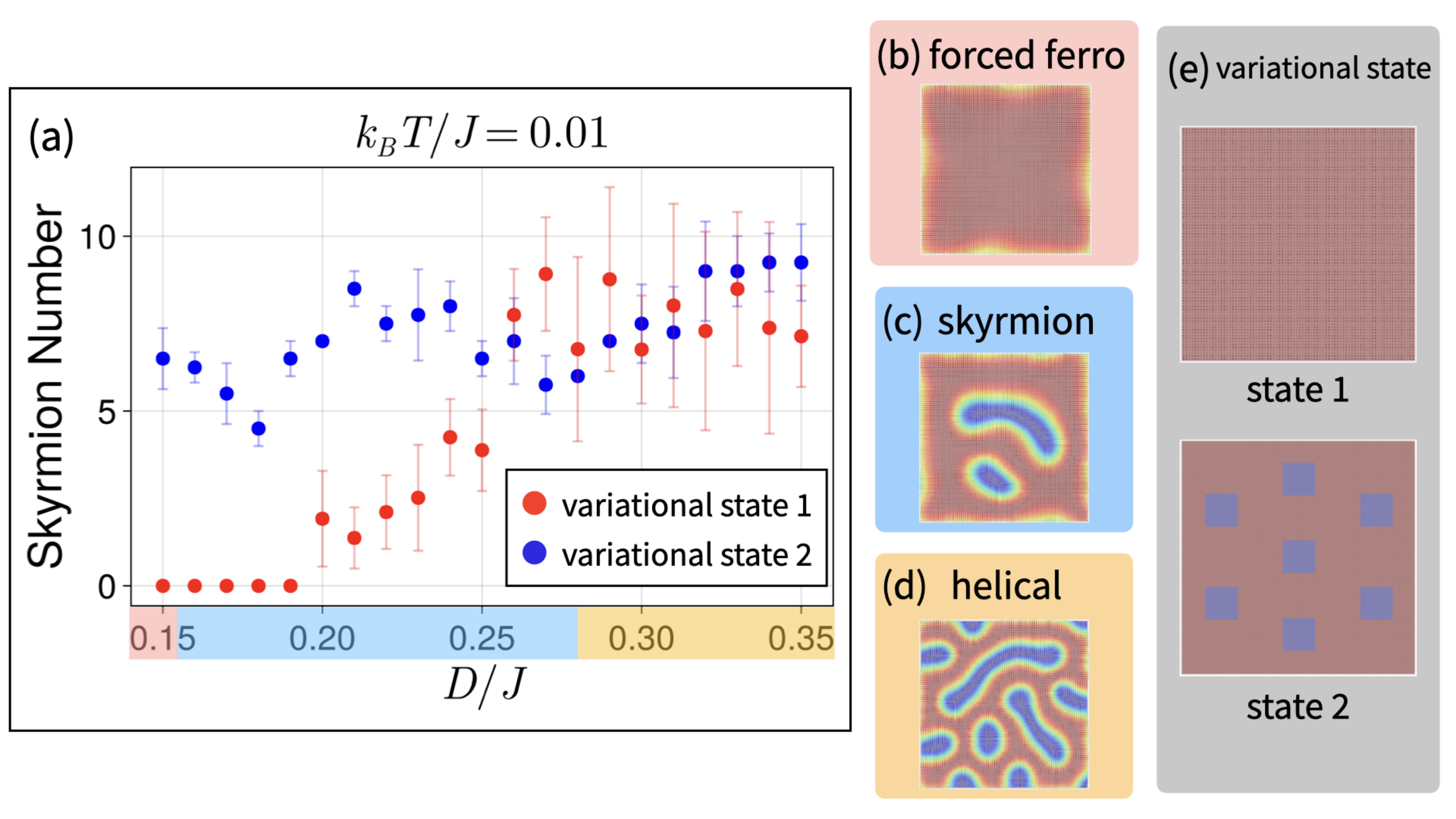}
  \caption{
  (a) $|\bm E|$ dependence of skyrmion number in the steady state at $t=t_f$ and $k_BT/J=0.001$. The horizontal axis is set to $D(\bm{E})/J \propto |\bm E|$ instead of the field strength $|\bm E|$. 
  Red and blue points are respectively the skyrmion numbers derived from the variational states 1 and 2 [see the explanation about panel (e)]. 
  Error bars stand for the standard deviation of the $N_s=100$ initial states. 
  Each color on the horizontal axis corresponds to each ground-state phase: Red, blue, and yellow, respectively, correspond to forced ferromagnetic, SkX, and helical phases. Panels (b), (c), and (d) are, respectively, typical spin configurations of 
  forced-ferromagnetic, SkX, and helical steady states at $t=t_f$. 
  (e) We start our time evolution from two kinds of variational states, State 1 and State 2. }
  \label{fig:SkX_average}
\end{figure}

\begin{figure}[t!]
  \centering
  \includegraphics[width=\linewidth]{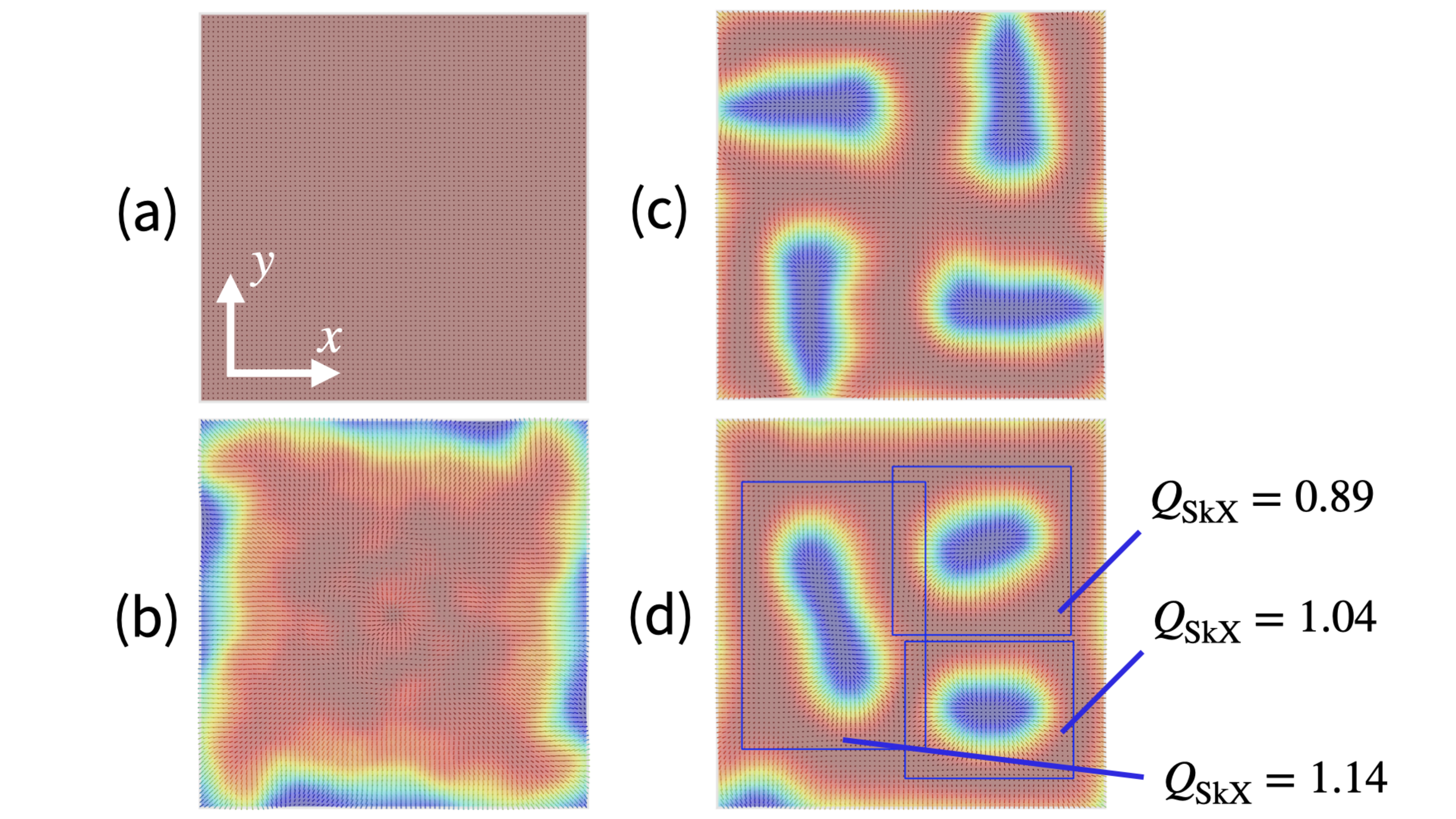}
  \caption{
  A typical time evolution of spins in the 2D spin model~\eqref{H_SkX_def} for $D=0.2J$. 
  The system size is $(N_x, N_y) = (80, 80)$.
  We show how the dc electric field induces the magnetic skyrmion by looking at snapshots of the time evolution of the LLG equation.
  (a) Forced ferromagnetic state at $t=0$. (b) State at $t=2t_0$ with $t_0 = 100\hbar/J$. (c) State at $t=1.5\times 10^3\hbar/J$, changing into the SkX state.
  (d) SkX state at $t=t_f= 7\times10^3\hbar/J$.
  The ML-based method enables the automatic detection of magnetic skyrmions, indicated by boxes.
  The skyrmion number within each box is also shown.
  Note that the skyrmion number is lower than 1 because of the thermal fluctuation at finite temperature and interactions between neighboring skyrmions.}
  \label{fig:SkX_dynamics}
\end{figure}

\begin{figure}[t!]
  \centering
\includegraphics[width=\linewidth]{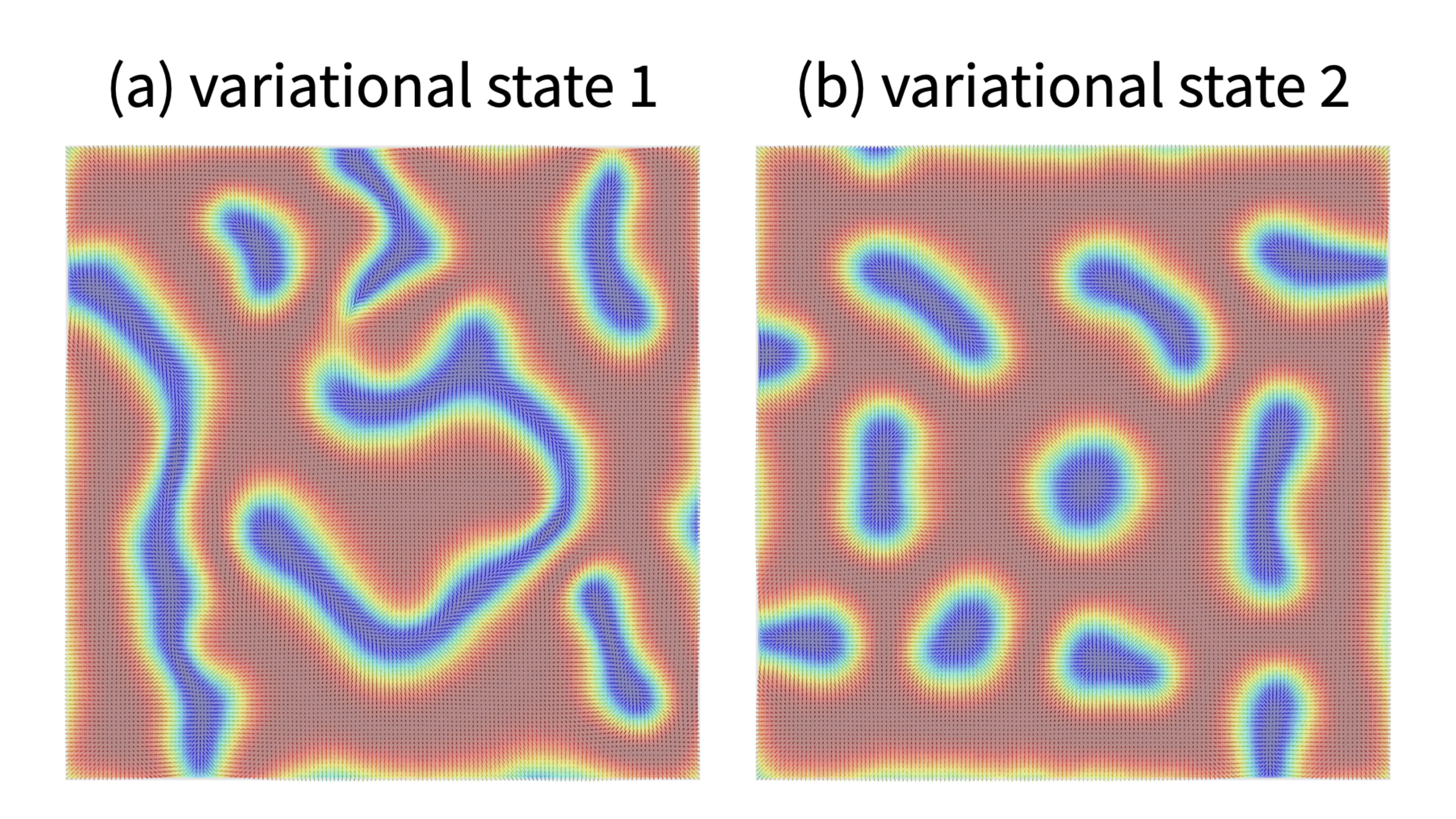}
  \caption{
  Comparison of two spin configuration snapshots at $t=t_f$ derived from the variational states 1 and 2. The system size is $(N_x,N_y)=(150,150)$.
  (a) Steady state at $t=t_f$ whose initial state is created from the ``initial state 1'' in Fig.~\ref{fig:SkX_average}~(e). The spin texture seems to have the nature of both the SkX and helix state. 
  (b) Steady state at $t=t_f$ whose initial state is the ``initial state 2'' in Fig.~\ref{fig:SkX_average}~(e). The spin texture seems to be the SkX state much more than that of the panel (a).
  }
  \label{fig:initial_states}
\end{figure}

\begin{figure}[t!]
  \centering
  \includegraphics[width=\linewidth]{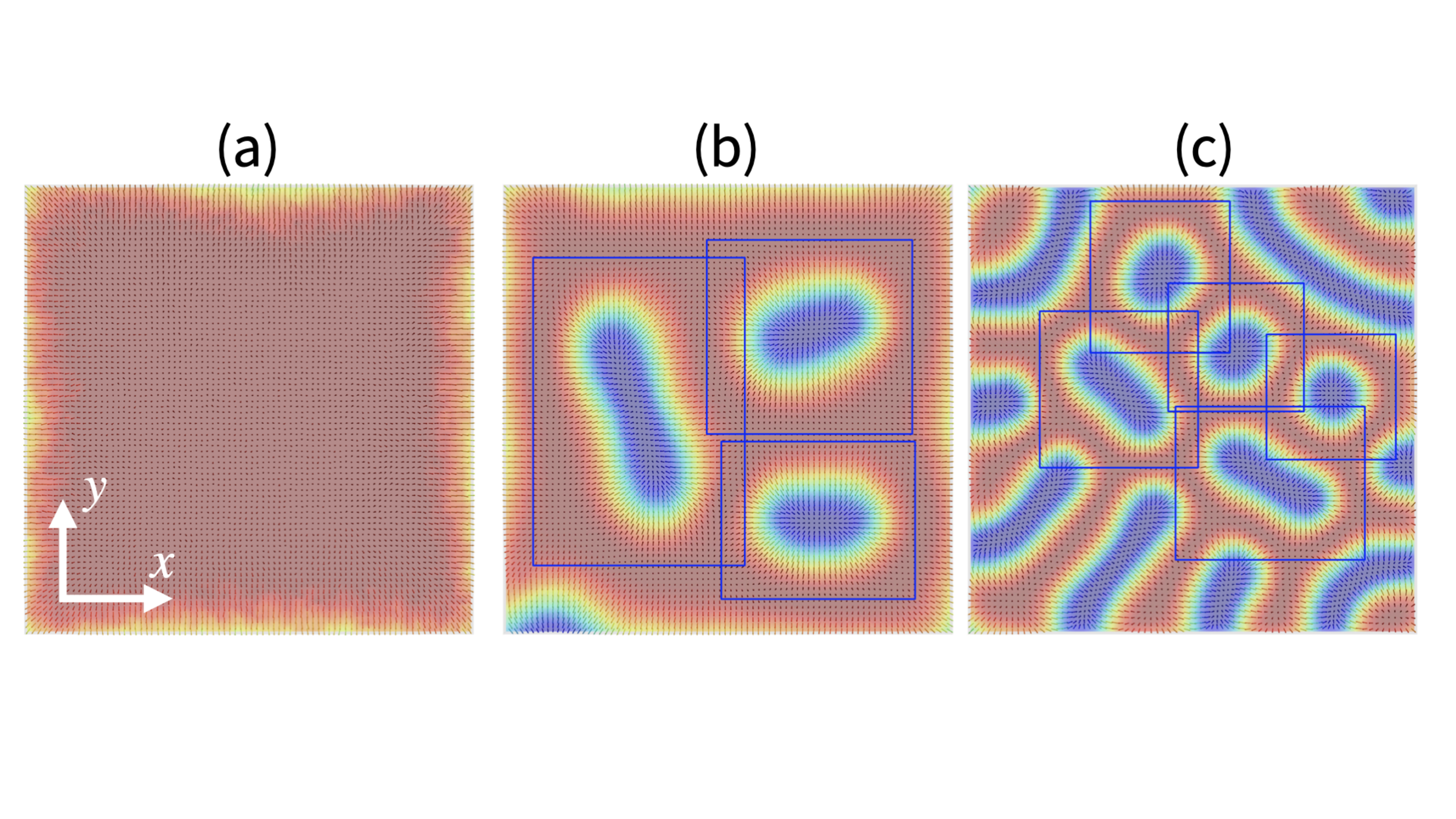}
  \caption{
   Typical snapshots of spin textures in the steady states at $t=t_f$ for (a) $D=0.15J$, (b) $D=0.2J$, and (c) $D=0.35J$. Each box in panels (b) and (c) means the size of the skyrmion surrounded by the box. 
   The system size is $(N_x, N_y) = (80, 80)$.
  }
    \label{fig:ref_2dim_spinTexture}
\end{figure}

Let us move on to the 2D case.
We show how the dc electric field drives the initial ferromagnetic state into the SkX state.
Figure~\ref{fig:SkX_dynamics} shows the spin dynamics that (a) the initial forced-ferromagnetic state eventually evolves into (d) the SkX state through two intermediate states (b) and (c).
Here, we set $J=1$, $\alpha=0.01$, $k_BT/J=10^{-3}$, and $B^z/J=0.018$.
The electric-field-induced DMI has the moderate strength 
$D(\bm E)/J = 0.24$.
We can expect the emergence of a SkX state at $t=t_f$ with this DMI strength, as indicated in the horizontal axis of  Fig.~\ref{fig:SkX_average}~(a).
We put the system on the square lattice with $N_x \times N_y$ sites ($N_\mu$ ($\mu=x,y$) is the site number along the $\mu$ direction).
We take $(N_x, N_y) = (150,150)$ for counting the skyrmion number in Fig.~\ref{fig:SkX_average}~(a) and take $(N_x, N_y) = (80, 80)$ to show snapshots of the time evolution (Figs.~\ref{fig:SkX_dynamics}, \ref{fig:initial_states}, and \ref{fig:ref_2dim_spinTexture}).
We adopt the open boundary condition along the $x$ and $y$ directions to emulate the thin film, the main platform where the SkX phase is experimentally realized.
Clock times of four states in Fig.~\ref{fig:SkX_dynamics} are (a) $t=0$, (b) $t=2t_0$, (c) $t=1.5\times10^3\hbar/J\gg t_0$, and $t=t_f = 7\times 10^3\hbar/J$.
The clock time $t_0$ of Fig.~\ref{fig:protocol}~(a) is set to $t_0=100\hbar/J$.
The panel (b) of Fig.~\ref{fig:SkX_dynamics} shows that the electric-field-induced DMI quickly twists the spin texture soon after the DMI is switched on.
Figure~\ref{fig:SkX_dynamics} clearly shows the time evolution of the forced-ferromagnetic state into the SkX state under the influence of the dc electric field through $D(\bm{E})/J$.

Figure~\ref{fig:SkX_average}~(a) shows the electric-field dependence of the topological number, i.e., the skyrmion number at $t=t_f$.
The horizontal axis is set to $D(\bm{E})/J\propto|\bm{E}|$ instead of $|\bm{E}|$.
We use two series of the time-evolution data, ``variational state 1'' and ``variational state 2'' in panel (a).
As it is obvious from their names, we obtain these two series by starting from two different states at $t=0$ [Fig.~\ref{fig:SkX_average}~(e)].
The ``variational state 1'' is the fully-polarized forced ferromagnetic state similarly to what we did in the 1D case. In this setup, we have an almost all-up state as the initial equilibrium state at $t=0$ like the 1D case. 
The ``variational state 2'' is a forced ferromagnetic state but with seven magnetic defects inside where the magnetic moments are inverted. We use this artificial state as the initial state at "$t=0$", namely, the initial state is non-equilibrium. 
The inversion of the magnetic moment gives zero topological number.
Still, the magnetic defects work as the seed of magnetic skyrmions. 
We hence expect that the spin dynamics starting from the ``variational state 2'' will more easily generate magnetic skyrmions, compared to the spin dynamics starting from the ``variational state 1''. 
It is difficult to experimentally prepare the "variational state 2", but we analyze the time evolution from the "variational state 2" as a comparison to the standard time evolution from the "variational state 1". For example, if the 2D Hamiltonian~\eqref{H_SkX_def} has the DMI with $D(E^z)/J\sim 0.2$, the corresponding ground state is expected to has six to eight skyrmions in the present $(150,150)$ size system because the linear size of single skyrmion is roughly estimated as $\sqrt{2/3}(2\pi)/(D(E^z)/J)\sim 50$ sites. 
Therefore, we introduce seven defects in the "variational state 2". 

Figure~\ref{fig:SkX_average}~(a) compares the skyrmion numbers of the steady states at $t=t_f$ that these different initial states finally reach after the long-time evolution.
We find a clear tendency that the skyrmion number associated with the "variational state 2" is larger than that with the "variational state 1".
This tendency is clearer for smaller $D/J$, when the steady state is expected to be either the forced ferromagnetic state or the SkX state.
Even after a long-time evolution, the steady state at $t=t_f$ remembers the initial state at $t=0$ more or less.
The red points in Fig.~\ref{fig:initial_states}~(a) show that in the case of the forced ferromagnetic initial state, the skyrmion number is suppressed even when $D(\bm{E})/J$ is large enough to give the SkX phase. This suppression would be due to the same finite-size effect as that discussed in the 1D case (see Sec.~\ref{sec:1d}).
Introducing the magnetically inverted defects to the initial state resolves this suppression and even enhances the skyrmion number, as shown in the blue points of Fig.~\ref{fig:initial_states}~(a). 
In addition to the finite-size effects, the skyrmion numbers generated from the variational states 1 and 2 also indicate the possibility that the final state at $t=t_f$ starting from the "variational state 1" cannot approach the true equilibrium state due to the energy barrier between topologically trivial and skyrmion states.

When the electric field is strong enough to give $D/J\geq 0.28$, the steady state is expected belong to the helical phase independent of the initial states.
The skyrmion number is nonzero in the helical phase because skyrmions emerge as isolate magnetic defects in this region. 
Figure~\ref{fig:ref_2dim_spinTexture} gives the comparison of steady states at $t=t_f$ for several different situations.
According to the ground-state phase diagram of Fig.~\ref{fig:phaseDiagram_CSL_SkX}~(b), we expect the forced ferromagnetic phase for $D/J=0.15$, the SkX phase for $D/J=0.2$, and the helical phase for $D/J=0.35$.
Indeed, the steady states of Figs.~\ref{fig:ref_2dim_spinTexture}~(a) and (b) are expected to be the forced-ferromagnetic state and the SkX state.
However, the spin texture of Fig.~\ref{fig:ref_2dim_spinTexture}~(c) does not look like helix even though this state is supposed to belong to the helical phase.
The finite-temperature effect is responsible for the apparent difference between this actual spin texture and the ideal helical structure.
At finite temperatures, thermal fluctuation disturbs the formation of the helical structure.
In other words, thermal fluctuation cuts the helix into pieces as shown in Fig.~\ref{fig:SkX_average}~(d).

Similarly to the 1D case, we point out the finite-temperature effects, especially the temperature range where magnetic skyrmions are maximally created. 
The finite temperature enables the system to overcome the energy barrier to create the TST, whereas it has the ability to destabilize the magnetic skyrmion. 
The creation of the magnetic skyrmion would be suppressed at temperatures below or above the proper moderate temperature range.
In addition to the chiral soliton, the magnetic skyrmion also fluctuates at finite temperatures.
At zero temperature, the magnetic skyrmion forms a lattice in the SkX phase.
Finite temperature liquefies the SkX and make the magnetic skyrmion fluctuate, i.e., a sort of skyrmion gas emerges (see a movie in Sec.~\ref{app:movie_2d}).

We can also find the thermal fluctuation in the helical phase.
The helical phase is supposed to show a "static" helix structure of magnetic moments.
However, the helix is broken in pieces [see Fig.~\ref{fig:SkX_average}~(d)].
The torn helices generate magnetic skyrmions as defects.
Therefore, there is no clear phase transition but a crossover between the SkX phase and the helical phase at finite temperatures.

\subsection{3D: Magnetic Hedgehog}\label{sec:3d}

\begin{figure}[t!]
  \centering
  \includegraphics[width=\linewidth]{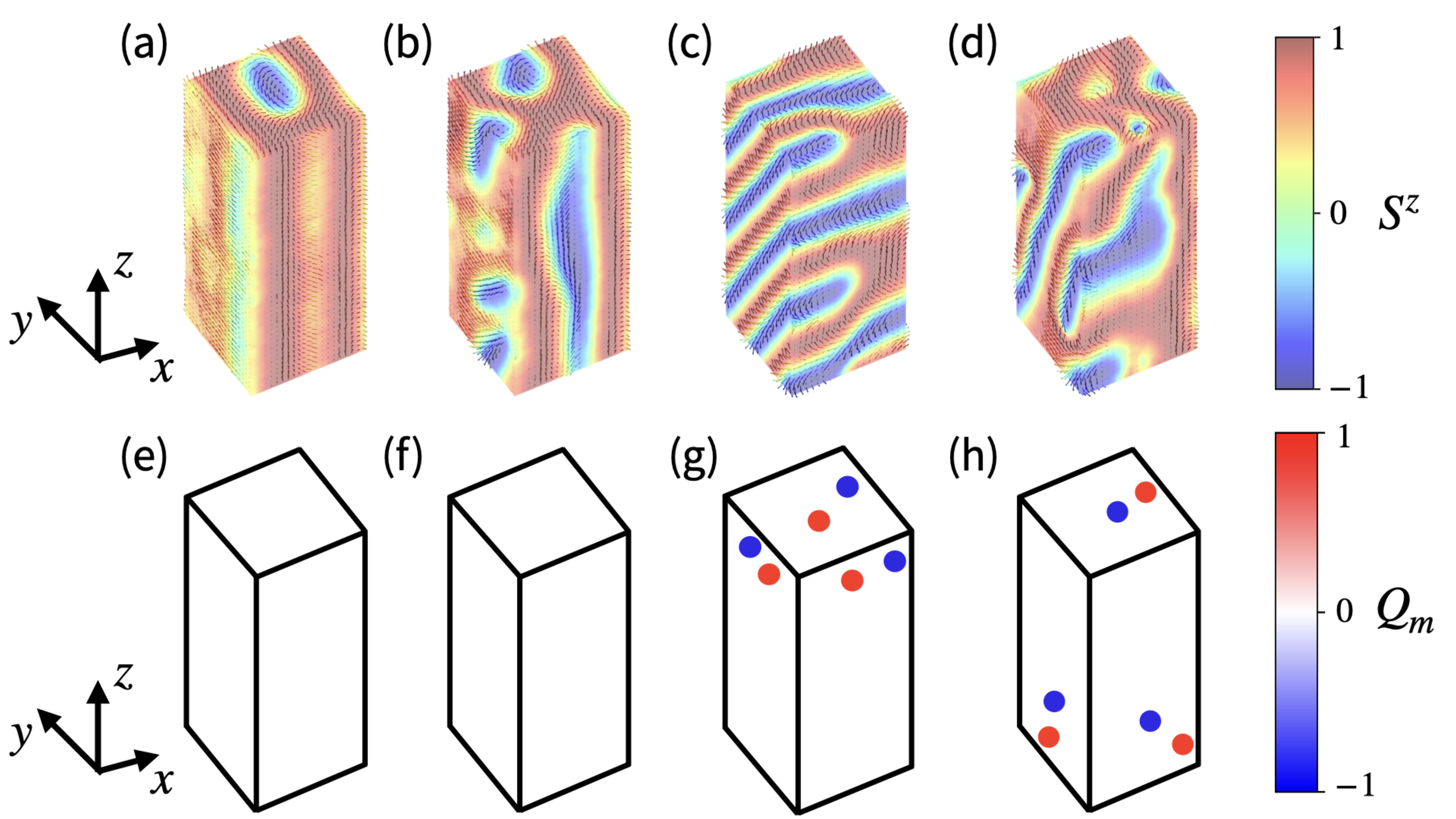}
  \caption{
  (a-d) Spin snapshots in a time evolution of the 3D model and (e-h) the location of the generated monopole charge $Q_m$. 
  (a) One-skyrmion-string state as the initial state at $t=0$.
  (b) State at $t=700\hbar/J$ (i.e., $t_0<t<t_1$) changing into the quadruple-$\bm q$ state but still with $Q_m=0$ in the entire region of the system.
  (c) State at $t=850\hbar/J$ (i.e., $t_1<t< t_f$) with nonzero monopole charges.
  (d) Steady state at $t=t_f=3000\hbar/J$ with nonzero monopole charges.
  (e) Monopole charge of the state of the panel (a), zero everywhere.
  (f) Monopole charge of the state of the panel (b), still zero everywhere.
  (g) Monopole charge of the state of the panel (c).
  We find the generation of monopole-antimonopole pairs.
  (h) Monopole charge of the steady state of the panel (d).
  We also observe monopole-antimonopole pairs.
  }
  \label{fig:hedgehog_dynamics}
\end{figure}

Finally, we discuss the 3D case.
In the 1D and 2D cases, we knew that the CSL state and the SkX state are realized as ground states of certain realistic Hamiltonians.
Unlike these cases, we have no such theoretical background for the magnetic hedgehog.
Nevertheless, we can find electric-field generation of the magnetic hedgehog state, using the Hamiltonian of \eqref{H_hdg_def} and the protocol of Fig.~\ref{fig:protocol}~(b) for application of the external electromagnetic fields.
Let us take a closer look at details of the spin dynamics.

We consider a brick of $N_x \times N_x \times N_z = 20 \times 20 \times 50$ sites, where we impose the open boundary conditions on the $x$ directions and the periodic boundary condition on the $y$ and $z$ directions. 
We take this boundary condition for a geometrical reason.
As Fig.~\ref{fig:condenser}~(b) shows, we apply the dc electric field along the $x$ direction, by using, say, a capacitor with plates parallel to the $y$-$z$ plane.
Then, it is reasonable to impose the open boundary condition instead of the periodic one to the $x$ direction.
We set parameters of the intrinsic Hamiltonian \eqref{eq:Hedgehog_ini} to $J=1$, $\alpha=0.01$, $B_0^z=0.12J$, and $D^{xy}=0.5J$ and those of the external one \eqref{eq:Hedgehog_ex} to $D^z=0.4J$, $B^x=0.01\times J$, and $B^y=0.02\times J$.
The two time scales $t_0$ and $t_1$ are chosen as $t_0 = 100\hbar/J$ and $t_1 = 800\hbar/J$ [see also Fig.~\ref{fig:protocol}~(b)]. We note that in this system size and the parameter setup, the initial equilibrium state at $t=0$ is expected to have only one skyrmion string along the $z$ direction, as shown in Fig.~\ref{fig:condenser}~(b).  
At $t=t_{-1}<0$, we start from the forced ferromagnetic state with a magnetic defect string of inverted magnetization (like the "variational state 2"), which is located in the center of the $x$-$y$ plane. This state with a defect string is expected to make a skyrmion string state quickly created. As expected, following this protocol, we obtain the equilibrium skyrmion string state at $t=0$.

\begin{figure}[t!]
  \centering
  \includegraphics[width=\linewidth]{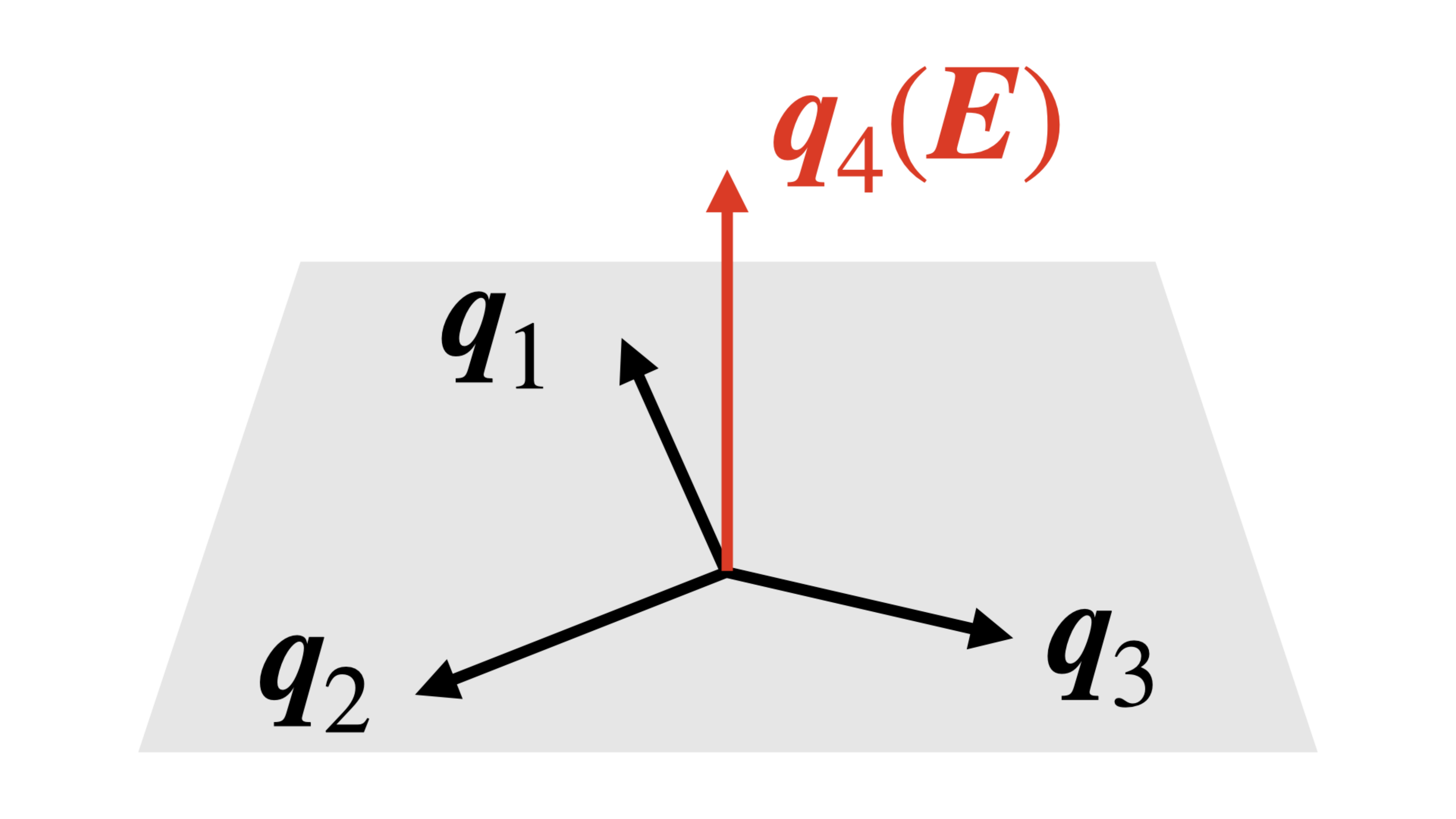}
  \caption{
  Schematic figure of $\bm q$ directions.
  The triple-$\bm q$ state has three $\bm q$ directions: $\bm q_1$, $\bm q_2$, and $\bm q_3$ on a plane.
  When the dc electric field is properly applied, another $\bm q$ vector, $\bm q_4(\bm E)$, grows.
  If $\bm q_4$ lives out of the plane spanned by $\bm q_i$ ($i=1,2,3$), we can expect the electric-field induction of magnetic hedgehog state~\cite{Yang_16,Okumura_20,Okumura_20_JPS,Shimizu_21,Shimizu_22}.
  }
  \label{fig:q_vectors}
\end{figure}

Figure~\ref{fig:hedgehog_dynamics} gives snapshots of the spin texture in a typical course of the time evolution from an initial skyrmion-string state to a final magnetic hedgehog state. 
As expected, the initial state in Figure~\ref{fig:hedgehog_dynamics} (a) has one skyrmion string, where the monopole charge obviously vanishes everywhere. 
This state has the triple-$\bm q$ state whose magnetic wavevectors $\bm q_i$ ($i=1,2,3$) live on the $x$-$y$ plane [Fig.~\ref{fig:q_vectors}].
We switch an electric field on at $t=t_0$.
For $t>t_0$, the system has the electric-field-induced DMI $D^y(E^x)$ in addition to the intrinsic one $D^{xy}$.
For $t_0<t<t_1$, the electric field is present but the transverse magnetic fields $B^x$ and $B^y$ in Eq.~\eqref{eq:Hedgehog_ex} are absent.
We can naively expect the emergence of hedgehogs with a monopole charge $Q_m \neq 0$ for $t_0<t<t_1$.
However, the present time evolution in Fig.~\ref{fig:hedgehog_dynamics} shows that 
the state at $t=700\hbar/J$ possesses no monopole charge anywhere. 
We further switch on the transverse magnetic fields within the $x$-$y$ plane for $t=t_1$. 
As we discussed earlier, the DMI $D^y(E^x)$ and the transverse field $\bm B$ are expected to cooperatively create hedgehogs. As expected, we confirm from Fig.~\ref{fig:hedgehog_dynamics} (g) and (h) that a quadruple-$\bm q$ state with a hedgehog-antihedgehog pair appears for $t>t_1$.
Interestingly, the monopole charge quickly becomes nonzero soon after $B^x$ and $B^y$ were switched on.
A state at $t=850\hbar J$ satisfying $0< t-t_1 \ll t_1$ shows a clear spiral structure [Fig.~\ref{fig:hedgehog_dynamics}~(c)] accompanied by nonzero monopole charges [Fig.~\ref{fig:hedgehog_dynamics}~(g)].
The generation of a monopole with $Q_m=1$ and an antimonopole $Q_m=-1$ occurs in a pair.
The magnetic hedgehog structure is being stabilized for $t>t_1$.
The system reaches the steady state at $t=t_f=3000\hbar/J$ [Fig.~\ref{fig:hedgehog_dynamics}~(d)] with nonzero monopole charges [Fig.~\ref{fig:hedgehog_dynamics}~(h)].
We here stress that there is a small but finite possibility that the dc electric field generates a magnetic hedgehog structure without the help of the transverse magnetic field.  

\begin{figure}[htbp]
  \centering
  \includegraphics[width=\linewidth]{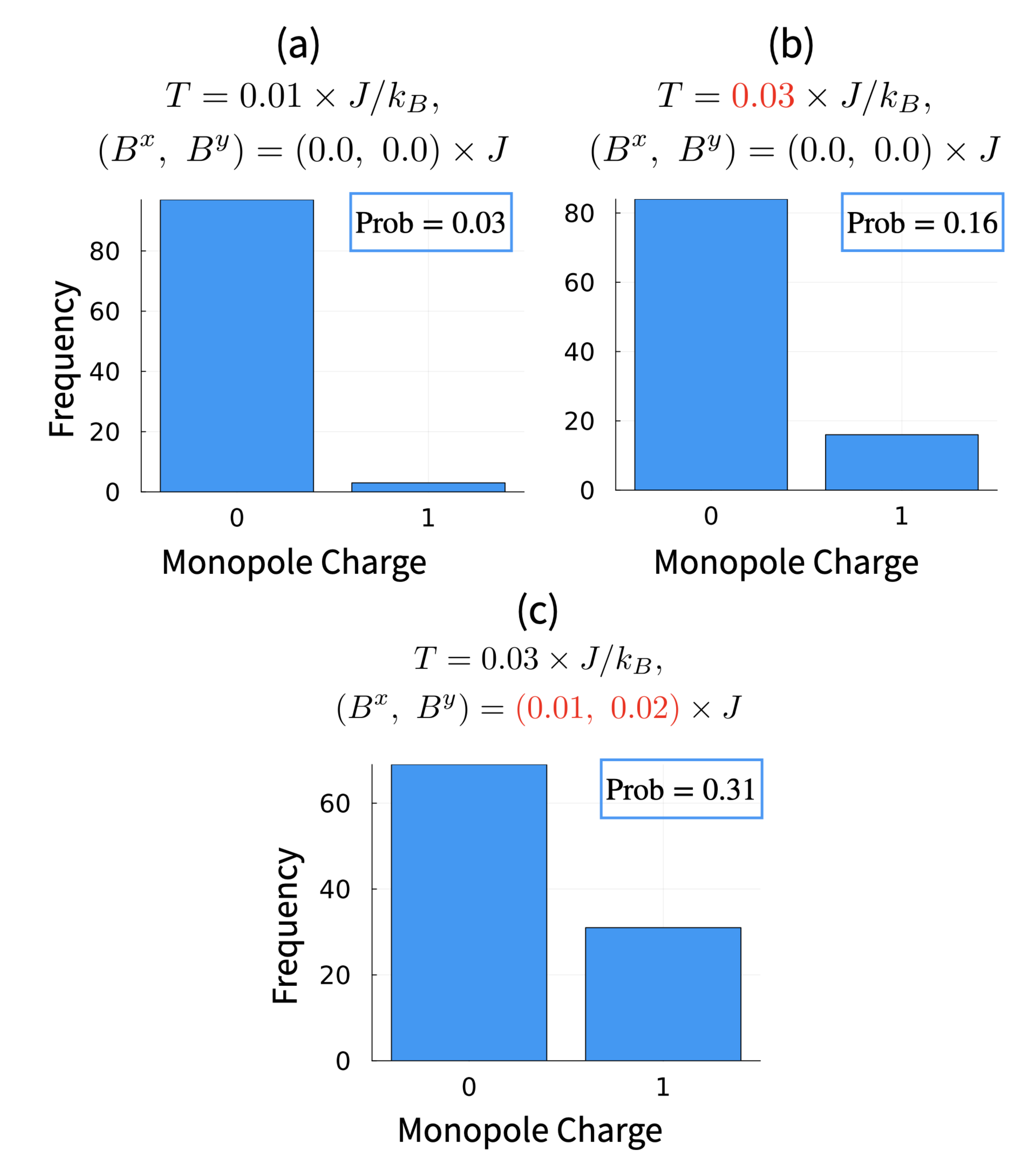}
  \caption{
  Probability of having nonzero monopole charge spin textures in the steady state at $t=t_f$ under the condition of a given temperature and strength of transverse magnetic field.
  The vertical axis refers to the frequency of event where the steady state has zero or nonzero monopole charge.
  The former is referred to as ``0'' and the latter as ``1'' in the horizontal axis.
  The pair of monopole ($Q_m=1$) and antimonopole ($Q_m=-1$) is counted as ``1'' in these panels.
  (a) $T=0.01J/k_B$ and $(B^x,B^y)=(0,0)$.
  The probability is lowest among the three situations compared here.
  (b) $T=0.03J/k_B$ and $(B^x,B^y)=(0,0)$.
  Increase of the temperature while the transverse magnetic field is off gives rise to the nonzero monopole charge event more frequently.
  (c) $T=0.03J/k_B$ and $(B^x,B^y)=(0.01J, 0.02J)$.
  The probability is highest among the three situations, implying the importance of the transverse magnetic field.
  }
  \label{fig:Hedgehog_stat}
\end{figure}

We observed that the spin dynamics driven by the Hamiltonian~\eqref{H_hdg_def} and the protocol of Fig.~\ref{fig:protocol}~(b) eventually generates the magnetic hedgehog state with monopole-antimonopole pairs and the hedgehog state is stabilized (i.e., it has a long lifetime).
Note that our model of Eq.~\eqref{H_hdg_def} contains only short-range interactions between local magnetic moments.
In addition, the model describes the magnetic Mott insulator without any itinerant electrons.
Namely, the model of Eq.~\eqref{H_hdg_def} offers a concrete example of the realistic spin model hosting hedgehog spin textures as its equilibrium state.

We emphasize that the transverse magnetic field $B^x$ and $B^y$ and temperature are critically important to enhance the probability of generating the magnetic hedgehog structure, shown in Fig.~\ref{fig:Hedgehog_stat}.
The figure represents the probability to find nonzero monopole charges at $t=t_f$.
When increasing the temperature in the absence of the transverse magnetic field, we find an increase in the probability of finding the nonzero monopole charge $Q_m=1$ [compare Figs.~\ref{fig:Hedgehog_stat}~(a) and (b)].
The importance of temperature is understood as follows.
Generally, four-spin interactions such as the biquadratic one $(\bm S_{\bm r} \cdot \bm S_{\bm r'})^2$ tend to stabilize a multiple-$\bm q$ magnetic order in the ground state at $T=0$.
Since our model~\eqref{H_hdg_def} contains two-spin interactions but no four-spin interactions, the ground state would not be the quadrupole-$\bm q$ ordered state with magnetic hedgehogs.
However, some previous studies indicate that in addition to the four-spin terms, multiple-$\bm q$ orders can be stabilized by an entropic effect due to finite temperatures~\cite{Reimers_pyrochlore_1991,Okubo_multiple-q_pyrochlore}. 
The magnetic hedgehog state is thus highly likely to be realized in our model only at finite temperatures.

When applying the transverse magnetic field with a fixed temperature, we also find the growth of the probability [compare Figs.~\ref{fig:Hedgehog_stat}~(a) and (c)].
As we have already discussed, the reason why this transverse magnetic field is introduced 
is that the competition between such a transverse field and the DMI is the key in the generation of the chiral soliton and the magnetic skyrmion, where the magnetic field is perpendicular to the direction of the DM vector.
The application of the transverse magnetic field in a proper temperature range thus makes the generation of the magnetic hedgehog state much easier.
Although it is an important problem to accurately determine the proper temperature range, the problem goes beyond the scope of this paper, and we keep it open for future investigations.

It is worth noting that the probability of finding the monopole does not reach unity (0.31 at highest in Fig.~\ref{fig:Hedgehog_stat}). 
This would be due to the fact that 
the system size we have simulated is only $20\times 20 \times 50$. 
This system size is large enough to discuss the generation of magnetic monopoles but still much smaller than the real material.
The probability will increase more as the system size becomes larger. 
Therefore, it is naturally expected that 
the probability of finding the monopole in real materials (sufficiently large systems) would be close to unity if the system is in the condition of Fig.~\ref{fig:Hedgehog_stat} (b) or (c). 
The numerical calculation in a larger 3D model is very tough in the sense of numerical cost: An extensive computation is necessary to observe the time evolution and to take the thermal average. 
As we have mentioned above, our calculation clearly shows the process of generating a hedgehog-antihedgehog pair (a monopole-antimonopole pair) 
by a dc electric field. Therefore, we will leave the issue of larger systems for the future.

\section{Summary}\label{sec:summary}

This paper numerically investigated the time evolution of the electron spins and the generation of TSTs in magnetic Mott insulators under the influence of dc electric fields.
Our numerical analysis is based on the generic framework of the many-body LLG equation.
The local spin $\bm S_{\bm r}$ feels the effective magnetic field $\bm B_{\mathrm{eff};\bm r} = \partial \mathcal H/\partial \bm S_{\bm r} + \bm \xi_{\bm r}$ determined by the many-body spin-spin Hamiltonian $\mathcal H$ and disturbed by the finite-temperature random field $\bm \xi_{\bm r}$.
Solving the microscopic LLG equation with a certain initial state, we obtain a final state so that the final state is a steady state or the same as the equilibrium state in the ideal situation.
This generic framework is explained in Sec.~\ref{sec:method}.

We have applied this numerical method to 1D, 2D, and 3D spin systems with short-range spin-spin interactions (see Sec.~\ref{sec:hamiltonian}) so that the final steady state possesses a TST; the CSL in the 1D case, the SkX in the 2D case, and the magnetic hedgehog in the 3D case. 
In Sec.~\ref{sec:results}, we have numerically confirmed that a strong enough dc electric field and the field-driven DMI eventually change a topologically trivial state into a non-trivial TST state.
We have detected the generation and number of TSTs by using the ML-based method in the 1D and 2D cases.
The spin texture, a spin configuration of a state at a fixed time, is processed as an image.
Based on machine learning of many spin-texture images, we can automatically detect TSTs. 
We have succeeded in counting the number of chiral solitons and magnetic skyrmions in the 1D and 2D models, respectively. 
On the other hand, in the 3D case, we have used the definition of the monopole charge~\eqref{Qm_hdg_def} to detect the monopole and antimonopole associated with the magnetic hedgehog texture.

We emphasize that in the 3D case, any parent Hamiltonian for a magnetic hedgehog state has never been known so far. 
Namely, no generic Hamiltonian with short-range interactions is yet available to realize the magnetic hedgehog state. 
However, following our strategy explained in Sec.~\ref{sec:hamiltonian}, we have succeeded in constructing a short-range interacting spin model hosting magnetic hedgehogs.

We have discussed the finite-size effect that suppresses the creation of TSTs, especially in the 1D and 2D cases. 
We argue that each TST occupies a finite volume, and hence the occupation number (or density) of TSTs in finite-size systems generally smaller than that in the thermodynamic limit. 

Temperature has a similar effect to the system size but in a more complicated way. 
Temperature plays a role in reducing the number of TSTs as it introduces the thermal fluctuation to the system. The thermal fluctuation effect is generally strong in low-dimensional systems. 
We have verified in Secs.~\ref{sec:1d} and \ref{sec:2d} that the thermal fluctuation makes 
the crystal-like TST states in the ground state (like CSL and SkX lattice) broken into a kind of TST gases or disordered TST. 
On the other hand, we should note that temperature is indispensable for turning the topologically trivial initial state into the TST state because temperature enables the system to overcome the energy barrier between these states.

Finally, we comment on other elements that can potentially increase the number of TSTs: Impurities and defects.
Real materials always have impurities more or less.
Defects can also be seen as a special case of nonmagnetic impurities.
It has recently been argued that these impurities of the system may increase the probability of emergence of TSTs~\cite{Iwasaki2013}. 
It will be important to discuss the impurity effect under the presence of dc electric fields: whether dc electric fields enhance the impurity effect or not.
However, this issue goes far beyond the scope of this paper. We leave it open for future studies.

\section*{Acknowledgments}

The authors are grateful to Shun Okumura for fruitful discussions.
M. S. is supported by the JSPS KAKENHI (Grants No. JP25K07198, No. 25H02112, No. JP22H05131, No. JP23H04576, No. JP25H01609 and No. JP25H01251), and JST, CREST Grant No. JPMJCR24R5, Japan. 
M. K. is supported by JST, the establishment of university fellowships towards the creation of science technology innovation (Grant No. JPMJFS2107), and by JST SPRING (Grant No. JPMJSP2109).

\appendix

\section{Critical DMI strength in 1D}\label{app:Dc}

As mentioned in Sec.~\ref{sec:1d}, the critical DMI strength that separates the forced-ferromagnetic and the CSL phases is known [Eq.~\eqref{critical_DMI_strength}] in the 1D model~\cite{Kishine_15}.
Let us call it $D^c$:
\begin{align}
    D^c = \frac{4}{\pi}\sqrt{B_zJ}.
    \label{eq:Dc_theory_app}
\end{align}

This appendix gives a brief review of the derivation of Eq.~\eqref{eq:Dc_theory_app} for the self-containedness of the paper. 
We assume that spins reside on the $S^x$-$S^z$ plane and ${\bm S}_j=S(\sin\phi_j,0,\cos\phi_j)$ at $T=0$ because the DM vector pointed to the $y$ direction in Eq.~\eqref{H_CSL_def}. 
The quantity $\phi_j$ denotes an angle formed by $\bm{S}_j$ and $\bm e_{z}$. 
As a result, the Hamiltonian \eqref{H_CSL_def} is rewritten as
\begin{align}
   \mathcal{H}_{\mathrm{CSL}}
    &= \sum_j \left[ -JS^2\cos\ab(\phi_j-\phi_{j+1})\right.\notag \\
    &\quad\left. - DS^2\sin\ab(\phi_j - \phi_{j+1}) -B^z S\cos\phi_j \right].
\end{align}
Here, we map this discrete model into a continuous model by replacing $ \sum_{j} \to a_0^{-1} \int_{0}^{Q_{\mathrm{CSL}}L} dx $ so that the spin chain is compatible with the spatial period $L$ of the CSL.
Recall that $Q_{\mathrm{CSL}}$ is an integer.
Taking this continuum limit, we can map the spin chain into a 1D chiral sine-Gordon (csG) model:
\begin{align}
  \mathcal{H}_{\mathrm{csG}}\ab[\phi]
  &\coloneqq
  \frac{\mathcal{H}_{\mathrm{CSL}}}{JS^2a_0}
  \notag \\
  & =\int_0^{NL}dx\ab[ \frac{1}{2} \ab(\frac{d\phi}{dx})^2 -Q_0\frac{d\phi}{dx} -m^2\cos\phi ] \label{eq:CsG_model},
\end{align}
where the variable $\phi(x)$ is the continuous limit of the angle $\phi_j$. 
We have introduced two parameters,
\begin{align}
    Q_0
    \coloneqq \frac{D}{a_0J},~
    m^2 
    \coloneqq \frac{B^z}{a_0^2JS}. \label{eq:CsG_model_parameter_def}
\end{align}
We normalize The Hamiltonian $\mathcal H_{\rm csG}$ by the factor $JS^2a_0$ to make it dimensionless.
The angle $\phi(x)$ is deemed a classical field and the only dynamical degree of freedom.
The field $\phi(x)$ is subject to an equation of motion,
\begin{align}
  \frac{d^2\phi(x)}{dx^2} = - m^2 \sin\phi. \label{CsG_eom}
\end{align}
Equation~\eqref{eq:CsG_model} and \eqref{CsG_eom} lead to a relation about the period $L$ of the CSL,
\begin{align}
  L = \frac{8E(\kappa) K\ab(\kappa)}{\pi Q_0}, \label{CsG_period}
\end{align}
where $K(\kappa)$ and $E(\kappa)$ are the complete elliptic integrals of the first kind and the second kind, respectively.
The parameter $\kappa$ is determined from an equation
\begin{align}
  \frac{\kappa}{E(\kappa)}
  = \frac{4m}{\pi Q_0}. \label{eq:kappa_E_B_DMI}
\end{align}As the parameter $\kappa$ approaches $1$, the complete elliptic integrals become $E(\kappa) \to 1$ and $K(\kappa) \to + \infty$, leading to the divergence of the spiral period $L$: $L \to + \infty$.
This divergence corresponds to the phase transition from the CSL phase to the forced ferromagnetic phase.
Taking the $\kappa \to 1$ limit in Eq.~\eqref{eq:kappa_E_B_DMI}, we obtain the critical DMI strength
\begin{align*}
    D^c = \frac{4}{\pi}\sqrt{B_zJ},
    \label{critical_DMI_strength_app}
\end{align*}
where we set $S=a_0=1$ as we did in the main text.
As explained in Sec.~\ref{sec:1d}, the numerically obtained $D^c$ is smaller than the analytical one \eqref{eq:Dc_theory_app}.

\begin{figure}[t!]
  \centering
  \includegraphics[width=0.9\linewidth]{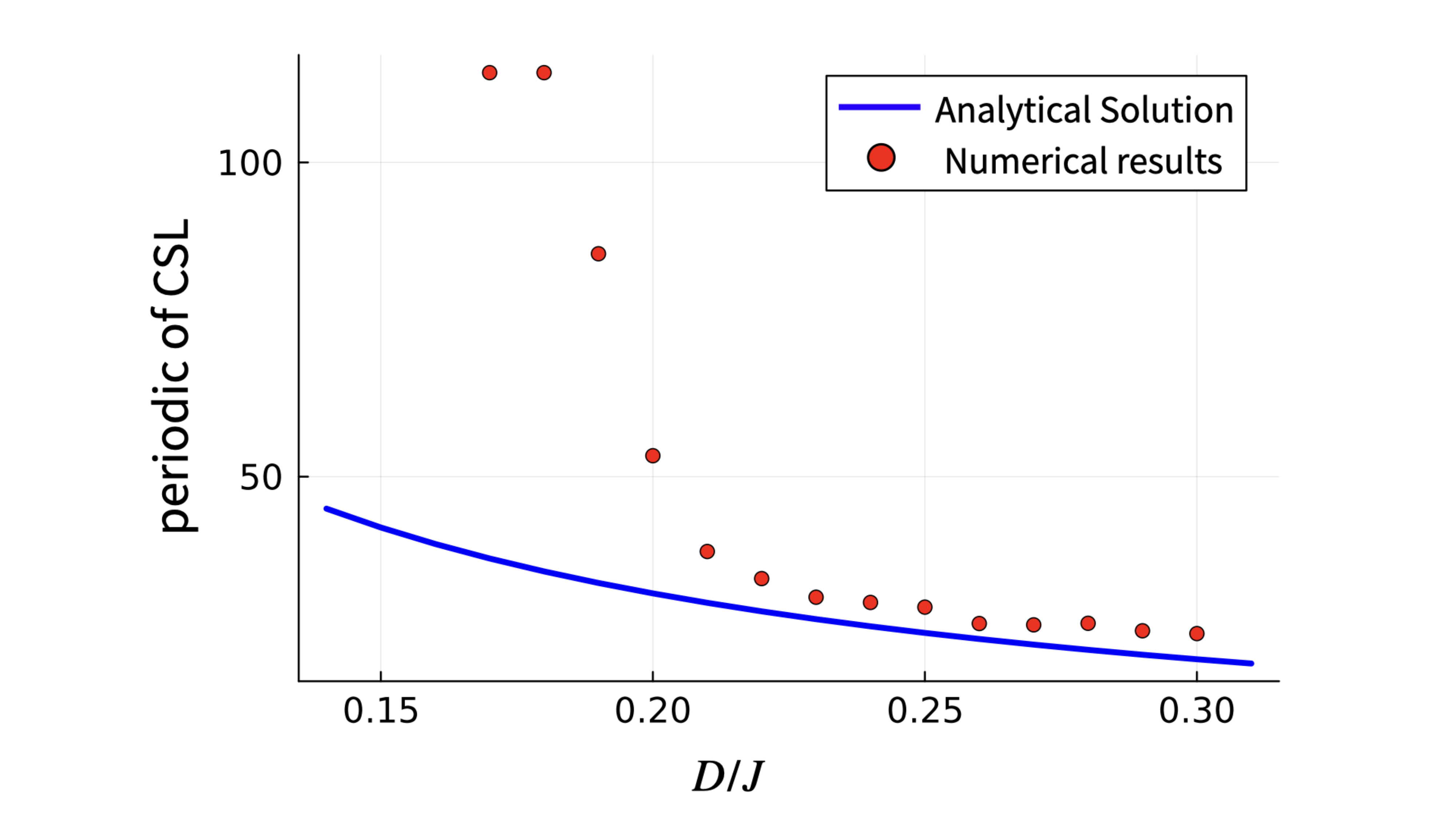}
  \caption{
  $D/J$ dependence of the spatial period of the CSL states.
  The solid curve is the analytical solution of Eq.~\eqref{CsG_period}.
  The red balls are numerical results, $N_x/\braket{Q_{\rm CSL}}$, estimated from the total soliton number $Q_{\rm CSL}$ at $t=t_f$. The bracket $\langle\cdots\rangle$ means the thermal average.   \label{periodDepDMI}
  }
\end{figure}

Let us compare the spatial periods $L$ computed from the continuous model~\eqref{eq:CsG_model} and the numerical method in Sec.~\ref{sec:1d}.
In the sine-Gordon model, $L$ of Eq.~\eqref{CsG_period} is determined by the parameter $\kappa$ obtained as a solution of Eq.~\eqref{eq:kappa_E_B_DMI}.
The parameter $\kappa$ is a function of $D/J$.
Accordingly, the period $L$ is also a function of $D/J$. 
Figure~\ref{periodDepDMI} gives a comparison of the analytical result of $L$ as a function of $D/J$ and the numerically obtained one $N_x/\braket{Q_{\rm CSL}}$ at $t=t_f$. 
Here, $\braket{Q_{\rm CSL}}$ is the total soliton number averaged over the canonical distribution of the random field [see Eqs.~\eqref{white_noise_av} and \eqref{white_noise_corr}]. 
In the range of moderate DMI ($D/J> 0.2$), the analytical and numerical results coincide with each other, while the numerically obtained period becomes far from the analytical one in the case of small DMI ($D/J< 0.2$). 
As we discussed in Sec.~\ref{sec:1d}, this deviation would be mainly attributed to the finite-size effect on the 1D model we have numerically treated.

\section{Soliton number distribution}

This section discusses the soliton-number distribution in the 1D model \eqref{H_CSL_def}.
As we explained in Sec.~\ref{sec:method}, the LLG equation contains the random field $\bm \xi_{\bm r}(t)$.
This random field is governed by the canonical distribution defined by Eqs.~\eqref{white_noise_av} and \eqref{white_noise_corr}.
Thus, the soliton number of the steady state at $t=t_f$ has a probabilistic nature.
To obtain the soliton number at a given temperature $T$, we need to take the average of the soliton number with respect to $N_s$ steady states.
Figure~\ref{fig:CSL_average}~(a) shows the $D/J$ dependence of such an averaged soliton number.
Here, we present the distribution of the soliton number before averaged.

\begin{figure}[t!]
    \centering
    \includegraphics[keepaspectratio, width=\linewidth]{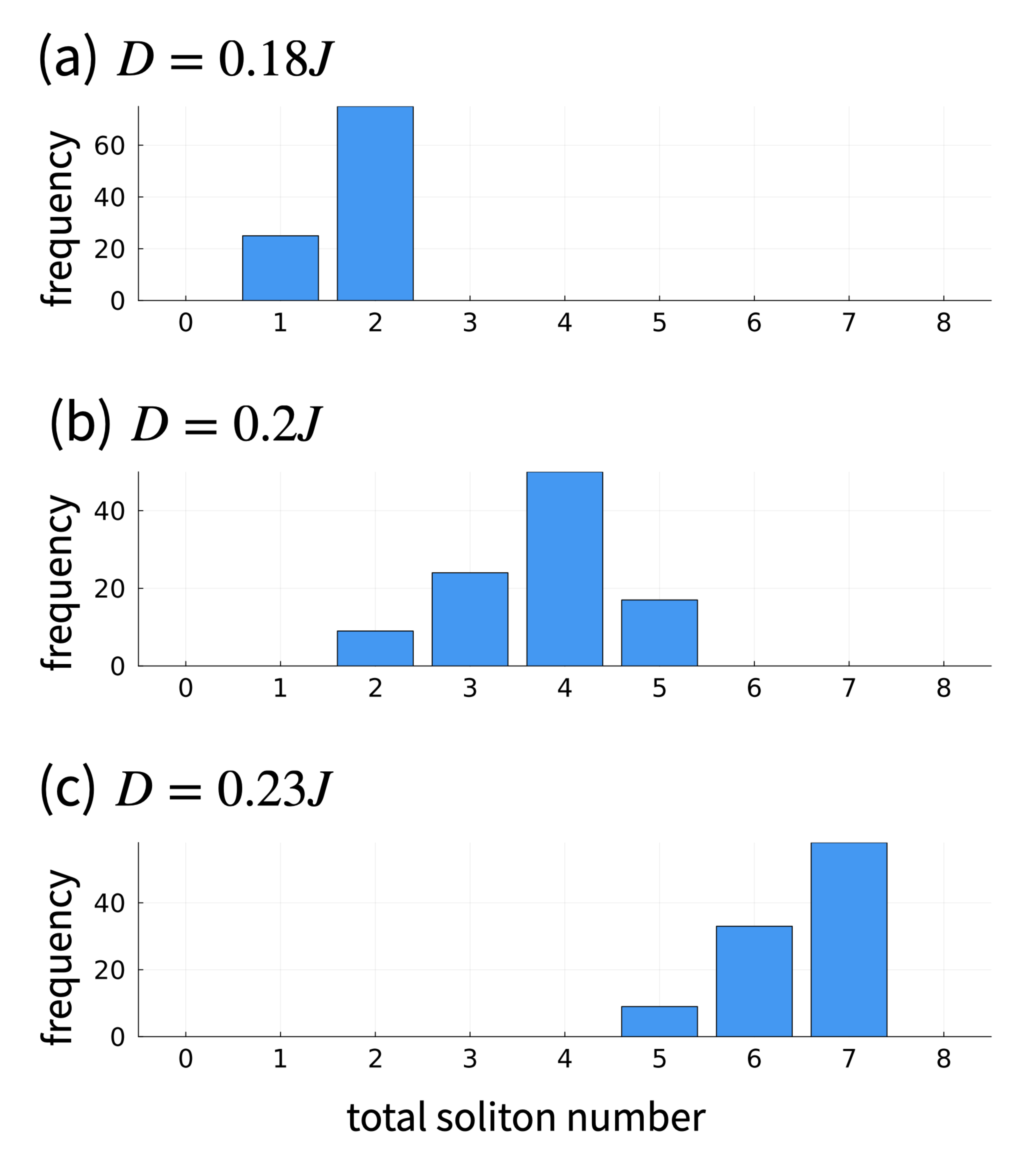}
    \caption{    
    Frequency distribution of total soliton number in the steady state at $t=t_f$ for (a) $D=0.18J$, (b) $D=0.2J$, and (c) $D=0.23J$.
    The mean value of the total soliton number in each panel is (a) 1.75, (b) 3.75, and (c) 6.40.
    }
    \label{SolitonNumber_Barplot}
\end{figure}

We use the following parameter set to discuss the electric-field dependence of the soliton number.
\begin{gather}
  N_x=200,~N_y=30,~J=1.0,~\alpha=0.01,\notag\\
  B^z=0.01J,~t_0=200~\hbar/J,~t_f=6000~\hbar/J,~N_s = 100, \label{param:Soliton}
\end{gather}
Here, $N_\mu$ ($\mu=x,y,z$) give the site number along the $\mu=x,y$, and $z$ directions.
The lattice spacing is set to unity. 
$N_s$ is the number of initial states required for the thermal average. 
Figure~\ref{SolitonNumber_Barplot} gives the soliton-number distribution.
The horizontal and vertical axes respectively refer to the total soliton number over the entire system and the frequency of each event that the steady state has the corresponding total soliton number.
It is clear that the steady state has a larger total soliton number as we increase the strength of the electric-field-induced DMI.

\section{Skyrmion number distribution}

\begin{figure}[t!]
    \centering
    \includegraphics[width=\linewidth]{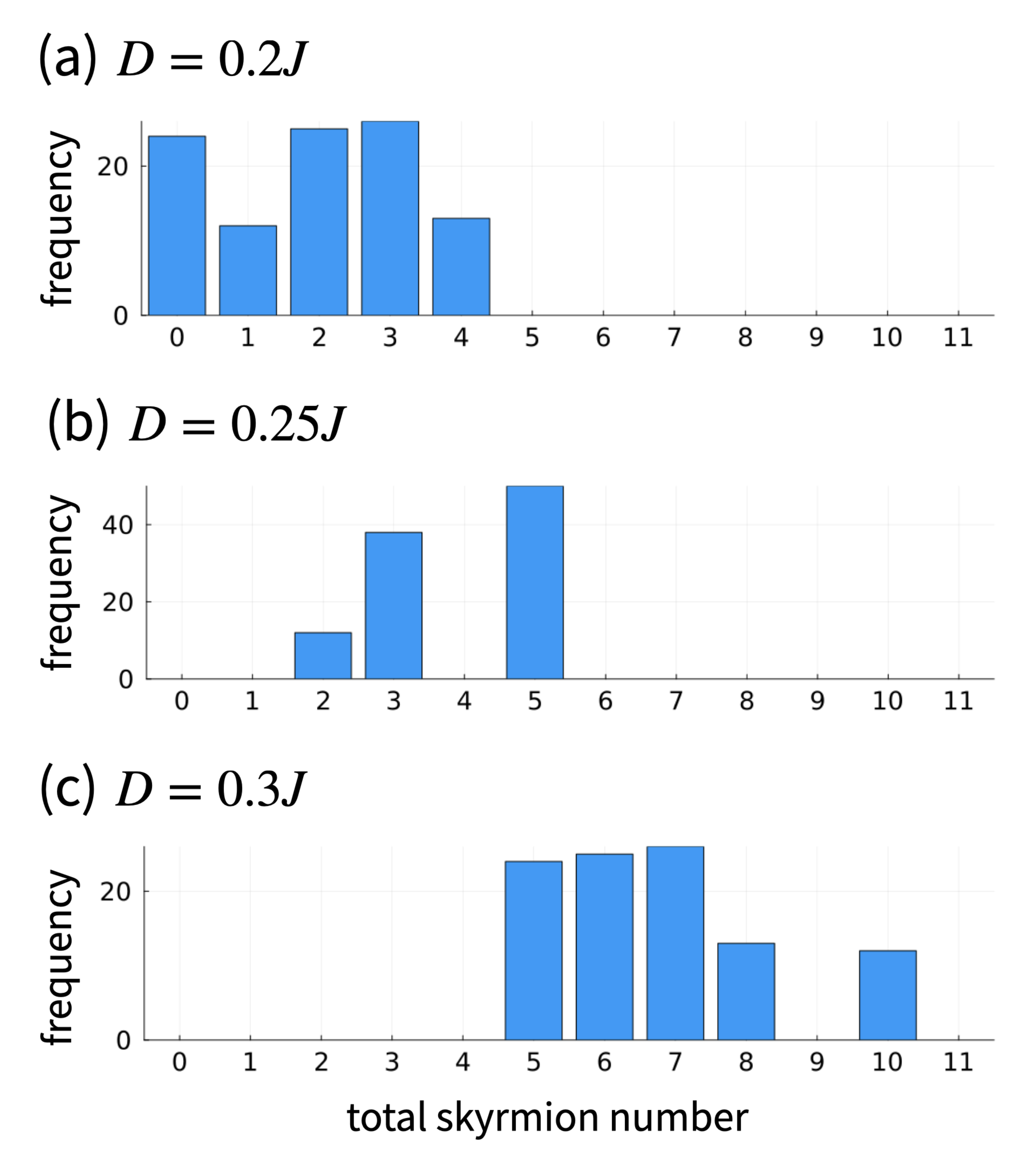}
    \caption{
    Frequency distribution of total skyrmion number for various DMI strength at temperature $k_{\rm B}T/J=0.01$ for (a) $D=0.2J$, $D=0.25J$, and $D=0.3J$.
    The skyrmion number is measured at $t=t_f$.
    Mean of the total soliton number in each panel is (a) 1.92, (b) 3.88, and (c) 6.76.
    }
    \label{SkyrmionNumber_Barplot_T0.01}
\end{figure}

\begin{figure}[t!]
    \centering
    \includegraphics[width=\linewidth]{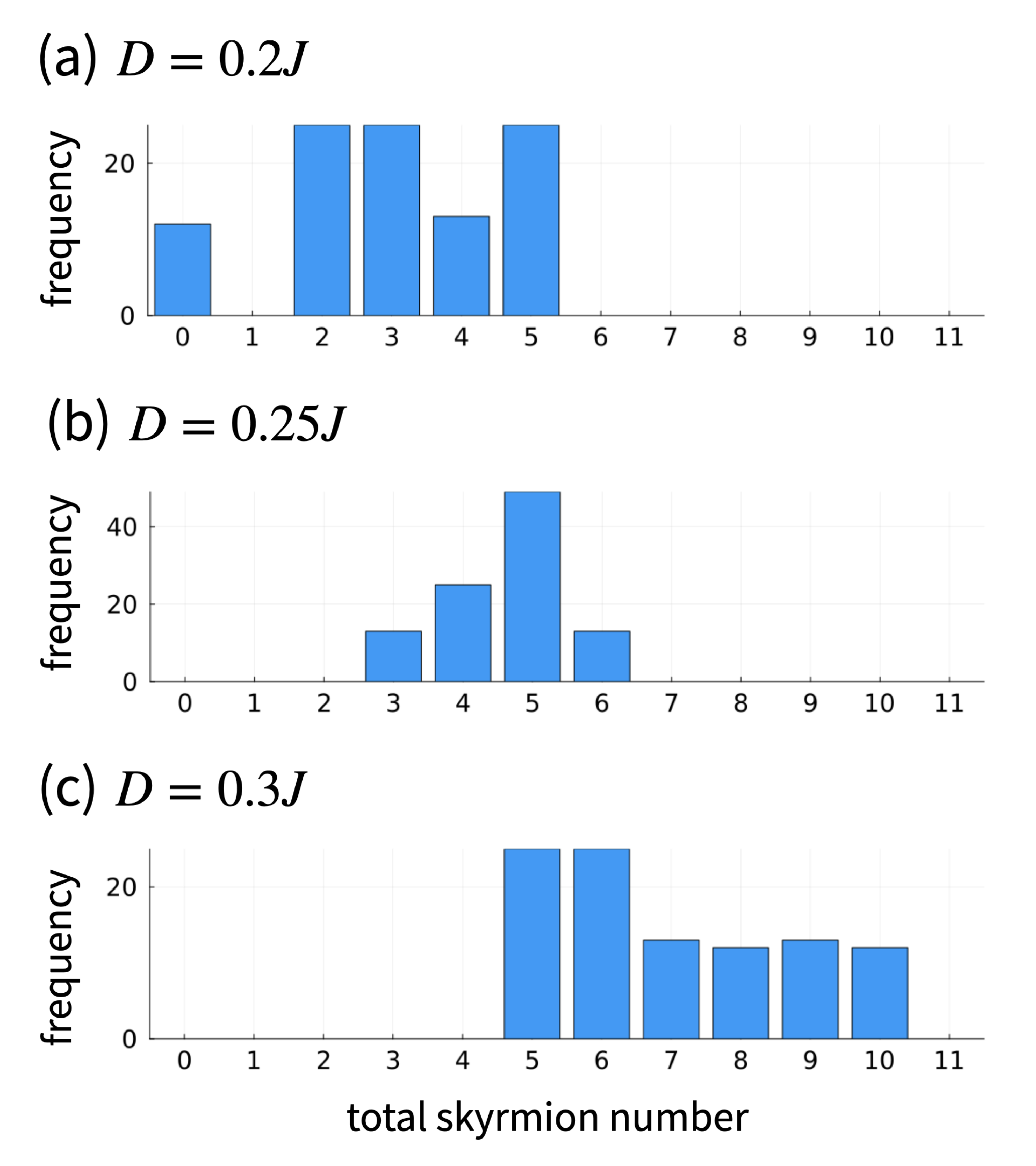}
    \caption{
    Frequency distribution of total skyrmion number for various DMI strength at temperature $k_{\rm B}T/J=0.001$ for (a) $D=0.2J$, $D=0.25J$, and $D=0.3J$.
    The skyrmion number is measured at $t=t_f$.
    Mean of the total soliton number in each panel is (a) 3.02, (b) 4.62 and (c) 6.99.
    }
    \label{SkyrmionNumber_Barplot_T0.001}
\end{figure}

Like the previous section, we here present the skyrmion number distribution of the steady state at $t=t_f$ in the 2D case.
We use the same parameters as those employed in Sec.~\ref{sec:2d}.
\begin{gather}
  N_x=N_y=150,~J=1.0,~\alpha=0.01,\notag\\
  B^z=0.018J,~t_0=200~\hbar/J,~t_f=6000~\hbar/J,~N_s = 100.
  \label{param:Skyrmion}
\end{gather}
We take the forced ferromagnetic state [Initial State 1 in Fig.~\ref{fig:SkX_average}~(e)] as the initial state of the spin dynamics simulation.
Figures~\ref{SkyrmionNumber_Barplot_T0.01} and \ref{SkyrmionNumber_Barplot_T0.001} show the frequency of the total skyrmion number of the steady state at temperatures $k_{\rm B}T/J=0.01$ and $0.001$, respectively, for several values of $D/J$.
The ground states for these values $D/J=0.2$, $0.22$, and $0.25$ belong to the SkX phase. 
One sees that the skyrmion number at $k_{\rm B}T/J=0.01$ is larger than that at $k_{\rm B}T/J=0.001$. 
This result indicates that the temperature $k_{\rm B}T/J=0.01$ is too high to maintain a long lifetime of skyrmions. We also find from the figures that the total skyrmion number is not a smooth function of $D/J$. This would be attributed to the finite-size effect.

\section{Movies}

Here we explain movies of the typical time evolution of spins in the 1D, 2D, and 3D cases. The movies are attached as Supplementary Materials. 

\subsection{1D: Chiral soliton lattice}\label{app:movie_1d}

In Movie-1~\cite{suppl}, we show the time evolution of the 1D chiral magnet \eqref{H_CSL_def} for $J=1$, $\alpha=0.01$, $k_BT/J=10^{-3}$, and $B^z/J=0.01$.
We consider the system of $N_x\times N_y = 200\times 30$ sites: $N_x = 200$ sites along the chain and $N_y = 30$ sites perpendicular to the chain.
We impose the open boundary condition on the $x$ and the $y$ directions. 
We follow the protocol of Fig.~\ref{fig:protocol}~(a) to apply the electric field.
We abruptly turn on the electric field at $t=t_0$ and keep it constant until $t=t_f$.
Here we set $t_0=100\hbar/J$ and $t_f =6000\hbar/J$.
For $t_0 < t \le t_f$, the system has the electric-field-induced DMI with the strength $D(E^z)/J=0.18$.
One can find how the forced ferromagnetic state turns into the CSL state in Movie-1.

The clock time ``$t$'' in the movie refers to the dimensionless one $tJ/\hbar$.
The color spectrum from red to blue depicts the local value of $S_{\bm r}^z$.
The red (blue) represents $S_{\bm r}^z=+1$ ($S_{\bm r}^z=-1$, respectively).
As we explained in Sec.~\ref{sec:ML}, we automatically detect the chiral soliton by using the ML-based method.
The soliton number is also calculated from its mathematical definition \eqref{Q_CSL_def}, where the interval $\ell$ is automatically determined by the ML-based method.
The ``id'' gives a label for each chiral soliton. 
There are two boxes below the plot of the spin texture. 
The first box (middle) shows the $t$ dependence of the external electric ($E^z(t)$) and magnetic ($B^z(t)$) fields.
The second box (bottom) shows the $t$ dependence of the total chiral-soliton numbers detected by the ML-based method and the topological number~\eqref{Q_CSL_def} in the detected domain.

\subsection{2D: Skyrmion lattice}
\label{app:movie_2d}

In Movie-2~\cite{suppl}.
we use the following parameters for the 2D chiral magnet~\eqref{H_SkX_def} on the square lattice: $J=1$, $\alpha=0.01$, $k_BT/J=10^{-3}$, and $B^z/J=0.01$.
We set the system size as $N_x \times N_y = 80\times 80$ sites.
We impose the open boundary condition on the $x$ and $y$ directions.
We follow the protocol of Fig.~\ref{fig:protocol}~(a) to apply the electric field, where $t_0 = 100\hbar/J$ and $t_f=6000\hbar/J$.
For $t_0<t\le t_f$, the system has the electric-field-induced DMI with the strength $D(E^z)/J=0.2$.
One can find how the forced ferromagnetic state turns into the SkX state in Movie 2.

Just like the 1D case, the clock time ``$t$'' in the movie refers to the dimensionless one $tJ/\hbar$.
The color spectrum is also the same as the 1D case.
We automatically detect the magnetic skyrmion by using the ML-based method.
The skyrmion number is also calculated from its mathematical definition \eqref{Q_SkX_def}, where the area $A$ is automatically determined by the ML-based method~\cite{Braun_12,Kim_20}.
The ``id'' gives a label for each magnetic skyrmion.
The two boxes below the plot of the spin texture are also similar to those in the 1D case.
The first box (middle) gives $E^z(t)$ and $B^z(t)$.
The second one (bottom) gives total magnetic-skyrmion numbers detected by the ML-based method and the topological skyrmion number~\eqref{Q_SkX_def} in the detected region.

\subsection{3D: Magnetic Hedgehog}
\label{app:movie_3d}

Movie-3 and Movie-4~\cite{suppl},
show typical time evolution of spins in the 3D model.
The model~\eqref{H_hdg_def} is defined on the cubic lattice.
We follow the protocol of Fig.~\ref{fig:protocol}~(b) to apply the electric field $\bm E(t) = E^y\bm e_y$ and the transverse magnetic field $\bm B(t) = B^x\bm e_x + B^y \bm e_y$.
The electric field is present for $t_0<t\le t_f$ and the transverse magnetic field is present for $t_1< t\le t_f$.
Note that $t_1>t_0$ holds.

We use the following parameters: $J=1,~\alpha=0.01,~B_0^z/J=0.1$, and $D^{xy}/J=0.4$.
We take the system size as $N_x \times N_y \times N_z = 20\times 20 \times 50$ sites.
We impose the periodic boundary condition on the $y$ and $z$ directions and the open boundary condition on the $x$ direction.
We set the time scales of Fig.~\ref{fig:protocol}~(b) to $t_0=700\hbar/J$, $t_1=1000\hbar/J$, and $t_f=3600\hbar/J$.
The electric-field-induced DMI with strength $D^z(E^y) = 0.4$ is present for $t_0<t\le t_f$.

One can find how the spin texture develops in Movie-3.
In addition, Movie-4 shows how magnetic hedgehog and antihedgehog (monopole and antimonopole) are created or annihilated. 
Movie-4 represents the monopole charge $Q_m$~\cite{Yang_16,Okumura_20,Okumura_20_JPS,Shimizu_21,Shimizu_22}, which is calculated from Eq.~\eqref{Qm_hdg_def}. Red and blue points respectively denote $Q_m=+1$ and $-1$. 
The position of the monopole charge is projected onto $x$-$y$, $y$-$z$, and $z$-$x$ planes.
Like the other two cases, the clock time ``$t$'' in Movie-3 and Movie-4 refers to the dimensionless time $tJ/\hbar$.

\clearpage


%

\end{document}